\documentclass[sigconf]{acmart}
\AtBeginDocument{%
  \providecommand\BibTeX{{%
    \normalfont B\kern-0.5em{\scshape i\kern-0.25em b}\kern-0.8em\TeX}}}

\copyrightyear{2021}
\acmYear{2021}
\setcopyright{rightsretained}
\acmConference[SC '21]{The International Conference for High Performance Computing, Networking, Storage and Analysis}{November 14--19, 2021}{St. Louis, MO, USA}
\acmBooktitle{The International Conference for High Performance Computing, Networking, Storage and Analysis (SC '21), November 14--19, 2021, St. Louis, MO, USA}
\acmDOI{10.1145/3458817.3480859}
\acmISBN{978-1-4503-8442-1/21/11}
\usepackage{graphicx}
\usepackage{stfloats}
\usepackage{subcaption}
\usepackage{indentfirst}
\usepackage{setspace}
\usepackage{url}
\usepackage{amsmath}

\usepackage{amssymb}
\usepackage{bm}
\usepackage{amsfonts}
\usepackage[ruled]{algorithm2e}
\usepackage{subfiles}
\usepackage{booktabs}
\usepackage{multirow}
\usepackage{makecell}
\usepackage{tabularx}
\usepackage{footmisc}
\usepackage{listings}
\usepackage{appendix}

\newcommand{\B}{\textcolor{black}}

\setcopyright{rightsretained}

\begin{document}

\title{Online Evolutionary Batch Size Orchestration for Scheduling Deep Learning Workloads in GPU Clusters}


\author{Zhengda Bian$^1$, Shenggui Li$^1$, Wei Wang$^2$, Yang You$^1$}
% \email{zbian@comp.nus.edu.sg}
% \email{sglee@nus.edu.sg}
% \email{wangwei.cs@gmail.com}
% \email{youy@comp.nus.edu.sg}
\affiliation{
    \institution{National University of Singapore$^1$, ByteDance$^2$}
    \country{Singapore}
}

\begin{abstract}

Efficient GPU resource scheduling is essential to maximize resource utilization and save training costs for the increasing amount of deep learning workloads in shared GPU clusters. Existing GPU schedulers largely rely on static policies to leverage the performance characteristics of deep learning jobs. However, they can hardly reach optimal efficiency due to the lack of elasticity. To address the problem, we propose ONES, an \textit{ONline Evolutionary Scheduler} for elastic batch size orchestration. ONES automatically manages the elasticity of each job based on the training batch size, so as to maximize GPU utilization and improve scheduling efficiency. It determines the batch size for each job through an online evolutionary search that can continuously optimize the scheduling decisions. We evaluate the effectiveness of ONES with 64 GPUs on TACC's Longhorn supercomputers. The results show that ONES can outperform the prior deep learning schedulers with a significantly shorter average job completion time.

\end{abstract}





\maketitle

\section{Introduction}
\label{sec:intro}

Deep Learning (DL) has gained significant success in the last decade.
Many efficient DL methods have been proposed, continuously renewing state-of-the-art results of machine learning (ML) tasks, such as computer vision \cite{verma2015large,szegedy2015going,he2016deep}, natural language processing \cite{radford2019language,devlin2018bert}, and speech recognition \cite{deepspeech2}, by taking advantage of the rising amount of data and growing scale of deep neural models.
However, training large-scale neural models can be extremely compute-intensive and time-consuming.
Therefore, people from both academia and industry are interested in accelerating DL jobs with distributed training on powerful GPU devices.
Recently, data parallelism techniques \cite{goyal2017accurate,you2017large,you2018imagenet,you2019large} have become dominant approaches to distributed training on huge datasets with large-scale GPU resources.
Data parallel training can fully utilize available GPUs by dividing a large minibatch to multiple devices, each of which executes an identical replica of the model and synchronizes with each other after each training step.
To meet the demand for GPU resources,
an increasing number of GPU clusters have been built, where users can share the expensive GPU resources.
In the meantime, resource management and job scheduling have become new challenges of training DL jobs in shared GPU clusters.

% Scheduling DL workloads in GPU clusters can be more intractable because of the specific performance of DL jobs, which largely depends on the resource allocation.
% According to the results in \cite{xiao2018gandiva} and \cite{jeon2019analysis}, increasing the number of distributed workers is not able to linearly accelerate a synchronous training job, since there will be an increasing communication overhead in synchronizing the model parameters.
Scheduling distributed DL jobs in a GPU cluster can be intractable \B{because of the specific performance of DL jobs w.r.t. resource allocation.}
Due to the communication overhead required by model synchronization and the affect of large minibatches on training convergence \cite{hoffer2017train,keskar2016large}, increasing the number of devices is not able to linearly accelerate a DL job.
% According to the results in \cite{xiao2018gandiva} and \cite{jeon2019analysis},
% different locality placement of the distributed workers can significantly affect the GPU utilization and training speed, because DL jobs are sensitive to the locality. 
Moreover, DL jobs may have severe performance interference between each other, especially when sharing the same GPU device, on which they contend for the GPU cores and memory \cite{jeon2019analysis}.
%It is therefore recommended to execute DL jobs on exclusive GPUs.
%It is therefore recommended to execute DL jobs on GPUs with non-preemptive mode.
%It is therefore recommended to execute DL jobs on GPUs exclusively.
\B{However, existing DL schedulers \cite{zhang2017slaq,peng2018optimus,xiao2018gandiva,oasis,gu2019tiresias} simply focus on the number of GPUs for the jobs, while they pay no attention to whether their batch sizes are appropriate.
Besides, these schedulers rely on greedy strategies to make decisions within a limited range of solutions.
As a result, the training efficiency of DL jobs are limited.}

Another issue arises from scheduling online workloads, which involve online job arrivals and completion.
DL jobs are iterative, so that it is hard to know the completion time in advance until they converge.
However, we can only make schedule decisions based on the information that already exists, while workloads in the cluster keep changing in real time.
As a result, we usually make stale and inefficient decisions for DL jobs.
A common solution of existing DL schedulers \cite{zhang2017slaq,peng2018optimus,oasis,gu2019tiresias} is to periodically update job status and reschedule them at a certain time interval.
Such schedulers are not able to make timely response to online workloads, as some jobs may be waiting between each rescheduling round, even when there already exist available resources.

\begin{figure}[t]
    \centering
    \includegraphics[width=\linewidth]{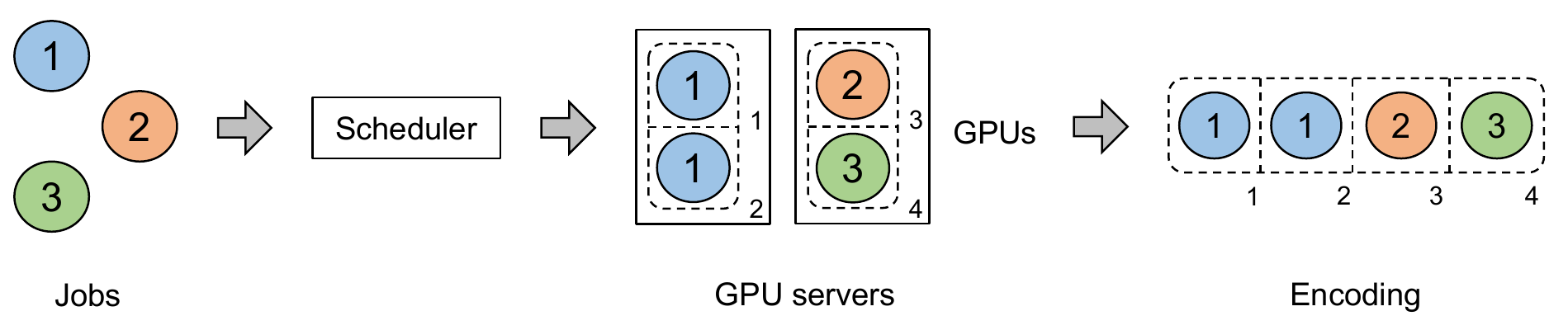}
    \caption{An example of scheduling GPUs for DL jobs.}
    \label{fig:encoding}
\end{figure}

To address \B{the above limitations in training and scheduling efficiency,}
we propose ONES, a new scheduler for DL workloads, which is based on a key idea that an elastic batch size orchestration for each DL job is essential to improve scheduling performance.
\B{The batch size of each job is \textit{elastic} because} we allow the scheduler to dynamically adjust it and find the optimal value to maximize training efficiency.
% because an core limitation in existing DL schedulers is that they simply focus on scheduling the number of GPUs for each job.
The first challenge to implement this idea 
%%ww: when you say "first challenge", I would expect "the second challenge...". But there is no such a sentence.
is to determine the batch size for each job.
%%ww: all-->every? 
We design an evolutionary search algorithm instead of greedy strategies to dynamically explore optimal decisions in an online fashion.
% The evolutionary search can be useful for improving scheduling performance in several ways.
% First, the algorithm maintains a group of solutions that are continuously optimized by fast evolution, which helps it find solutions that are close to optimum in a short time.
% Moreover, as shown in \autoref{fig:encoding}, scheduling GPUs can be easily encoded into genomes (i.e. the basic form of solutions in evolutionary search) so that the evolutionary search algorithm can be directly applied to the scheduling problem.
% Besides, the algorithm can be efficient to respond to online workloads by capturing real-time job status when the solutions are evolving.
% To select more suitable solutions from the candidates, the algorithm conducts online prediction on expected remaining job length for each job based on the training information of completed jobs, since the distribution of job lengths may be relatively stable.
\B{The algorithm maintains a group of solutions (i.e. population) that continuously evolve towards higher performance.
Hence compared with existing scheduling strategies, evolutionary search has the following advantages to obtain more efficient decisions.
First, the algorithm works with a population that helps the scheduler have more chance to make high-quality decisions.
Second, the algorithm is efficient to deal with online workloads because it continuously updates its solutions according to real-time job status.
Especially, ONES builds an online prediction model to obtain possible distributions of job lengths throughout training, from which the algorithm can sample solutions with shorter estimated average job completion time.
Besides, as shown in \autoref{fig:encoding}, scheduling GPUs can be easily encoded into genomes (i.e. the basic form of solutions in evolutionary search) so that the algorithm can be directly applied to the scheduling problem.
}
The second challenge is to execute scheduling decisions in an efficient way.
\B{We design an elastic batch size scaling mechanism, which minimizes the cost to execute changes in the batch size and number of GPUs of DL jobs that are induced by new scheduling decisions.
The mechanism uses a centralized scheduler to make scheduling decisions and manage batch sizes, and sends messages to specific nodes to execute scaling at only necessary workers.
Unlike common scaling approaches based on model checkpointing, ONES can automatically re-configure the batch size as well as number of GPUs through NCCL communication \cite{nccl} at the end of a training step without stopping training.
Thus, its cost is almost invisible to the original job.
}
% Thus, we design the batch size scaling mechanism that can be almost invisible to the original training, as it is able to adjust the batch size as well as the number of workers at the end of a training step without stopping the training process.
Besides, for the purpose of efficient utilization of limited GPU resources, ONES follows a set of scaling policies to guarantee the training performance of each job as well as the scheduling efficiency.
% Lastly, we implement our algorithms in a distributed system with centralized scheduler that runs each job by sending messages to specific nodes to execute the workers.

In summary, we make the following contributions.
\begin{itemize}
    \item We propose ONES, an \textit{online evolutionary scheduler} that is the first scheduler to schedule batch sizes for DL jobs to maximize resource utilization.
    % the training performance with \textit{elastic batch size scaling} (\S \ref{sec:scaling}) and searches for the optimal scheduling decision in an online fashion (\S \ref{sec:online}).
    \item We design an online scheduling algorithm based on evolutionary search.
    %%ww: utilization -->length?
    % To minimize JCT for DL jobs, we design the \textit{smallest remaining utilization first} algorithm (SRUF) that leverages the historical information of completed jobs to conduct  (\S \ref{sec:sruf}).
    %%ww: --> different to
    % Different to greedy strategies, the algorithm selects the best scheduling solutions by using probability sampling to filter the expected higher-quality candidates.
    Different to greedy strategies, our algorithm dynamically selects a superior population of solutions according to jobs' online performance prediction derived from their real-time status.
    \item We design elastic batch size scaling to execute adjustment of resource allocation with almost invisible cost, which does not need to stop the original training process.
    \item ONES is efficient in minimizing job completion time (JCT). The evaluations on a testbed of TACC's Longhorn GPU nodes \cite{longhorn} and trace-driven workloads show significant improvements over state-of-the-art DL schedulers.
\end{itemize}

\section{Background and Motivation}
\label{sec:background}

\subsection{Existing schedulers for deep learning}

Existing DL schedulers \B{have proposed} a variety of strategies that optimize specific performance of DL jobs based on greedy heuristics.
For example, SLAQ \cite{zhang2017slaq}, Optimus \cite{peng2018optimus} and OASiS \cite{oasis} aim to find the optimal job prioritization via performance analysis.
Tiresias \cite{gu2019tiresias} maintains a Multi-Level Feedback Queue based on the Least Attained Service policy to reduce queuing time.
Some other schedulers \cite{xiao2018gandiva, jeon2019analysis} focus on optimizing job locality and placement to minimize the affect of communication overhead.
However, the efficiency of existing scheduling algorithms still confronts the following limitations. 

\textbf{Lack of elasticity:}
Elastic job size can help the scheduler improve the cluster efficiency, as it gives the opportunity to increase resource utilization and reduce fragmentation.
Unfortunately, most existing schedulers are not aware of the elasticity and execute DL jobs with fixed amount of resources that are requested by users.
However, a fixed resource configuration can be inefficient in many aspects.
On one hand, allocating resources to DL training jobs has to follow the gang scheduling policy \cite{xiao2018gandiva}, which leads to longer queuing delay.
In gang scheduling \cite{papazachos2010performance}, all the workers of a job need to become active together, so that as long as their resource requirements cannot be fulfilled simultaneously, the allocation will fail.
Therefore, with aggressive configurations, many jobs need to wait for the available resources in the cluster until their requirements can be fulfilled, even though they are able to run at a smaller scale.
On the other hand, using conservative configurations results in low resource utilization and intolerable training time.
For example, it is reported \cite{goyal2017accurate} that training ResNet-50 \cite{he2016deep} on the ImageNet dataset \cite{imagenet} takes 29 hours to reach the state-of-the-art accuracy, with a batch size of 256 images on 8 Tesla P100 GPUs.
However, with a larger batch size of 8192, using 256 GPUs, the training time can be reduced to less than one hour.
Unfortunately, users often submit inappropriate configurations when they are not familiar with the low-level system,
% As a result, the jobs cannot benefit from any extra resource when it is available, unless a cluster manager manually re-configures the allocation, or the user restarts the job with a better configuration.
% This may 
which leads to an obvious resource under-utilization.

% Unfortunately, it is not easy to determine the amount of GPU resources for a job, because more workers may not necessarily improve the training speed due to higher communication overhead.
% As a result, users may waste great efforts in trial and error for an optimal configuration.

There are DL schedulers trying to improve the elasticity by dynamically adjusting the the number of GPUs for each job \cite{xiao2018gandiva,zhang2017slaq,peng2018optimus,LinYC19}.
However, the degree of elasticity in these schedulers is not efficient enough.
% They still 
They only focus on scheduling the number of GPUs and treat DL jobs as black-boxes, but are not aware to take a step forward to schedule the size of minibatches at the same time.
Consequently, 
inappropriate combinations of the number of GPU and batch size reduce the efficiency of distributed training, as
increasing the number of workers leads to increase in the communication overhead, so that the utilization of each single GPU will be decreased.

\textbf{Suboptimal scheduling quality:}
Existing resource schedulers focus on optimizing resource utilization through static greedy algorithms, such as Shortest Job First (SJF) policy \cite{tetris}, Least Attained Service (LAS) policy \cite{gu2019tiresias}, or time-sharing-based slicing strategy \cite{xiao2018gandiva}.
Unfortunately, such scheduling strategies still confront efficiency issues.
First, resource scheduling is an NP-hard optimization problem, so that the greedy solutions may often lead to local minima that is far from the optimal performance.
Second, existing schedulers consider DL jobs as black-boxes, not aware of the significant role of batch sizes in improving training efficiency.
Consequently, regardless of how the schedulers allocate available resources, the scheduling performance is largely limited.

% Theoretically, the \textit{shortest remaining processing time} (SRPT) is the optimal algorithm in minimizing average JCT \cite{grosof2018srpt}.
Besides, some DL schedulers \cite{zhang2017slaq,peng2018optimus} minimize average completion time by predicting the remaining length of each job.
However, predicting accurate job length can be difficult in practice and not scalable for all kinds of DL workloads.
On one hand, many DL jobs do not have smooth loss curves or validation accuracy curves (e.g. a reward function in RL and a loss with cyclical learning rate \cite{smith2017cyclical}).
On the other hand , not all DL jobs can end normally, as some jobs are manually killed, some are early-stopped, some crashed due to errors, etc.

% Without the information of job lengths, current resource schedulers for DL workloads basically use strategies based on simple heuristics.
% For example, Gandiva \cite{xiao2018gandiva} attempts to improve cluster utilization based on its packing strategies.
% Tiresias \cite{gu2019tiresias} attempts to reduce JCT with the least attained service (LAS) policy. This policy helps to reduce the delay and guarantees fairness, but may not be optimal for minimizing JCT. The result presented in the paper also shows that SRPT still outperforms Tiresias.
% Although these scheduler can improve JCT to some extent, they are still far from an optimal solution.

\textbf{Inefficient online scheduling:}
To make online scheduling decisions,
% Some existing schedulers (e.g. SLAQ \cite{zhang2017slaq}, Optimus \cite{peng2018optimus}, and OASiS \cite{oasis}) periodically update job status and reschedule them based on the analysis of job performance, such as predicting training progress or remaining processing time.
% However, such prediction is hard to be accurate, because practical training is not guaranteed to present smooth loss and accuracy curves.
% Besides, the schedulers use fixed scheduling intervals, so that they may not be able to make immediate response to changes in workloads.
% Some other schedulers (e.g. Gandiva \cite{xiao2018gandiva} and Tiresias \cite{gu2019tiresias}) continuously monitor the jobs and adopt event-driven strategies to handle job arrival and completion.
% They allocate GPUs to DL jobs based on fixed policies without any prediction about the jobs.
% Since the policies are basically not able to take advantage of the workload information, they can hardly be efficient all the time, and may lead to higher average Job Completion Time (JCT).
In order to make online scheduling decisions, most of existing schedulers (e.g. SLAQ \cite{zhang2017slaq}, Optimus \cite{peng2018optimus}, Tiresias \cite{gu2019tiresias}, and OASiS \cite{oasis}) periodically update job status and reschedule them based on the analysis of current job performance.
They basically use fixed scheduling intervals, so that they may not be able to make immediate response to changes in the workloads.
For example, Optimus's interval between 2 rounds of rescheduling is 10 minutes, and that of OASiS is 1 hour.
At each round, all the jobs will be paused to wait for the next scheduling decisions.
As a result, there will also be a nontrivial waste of resources because the GPUs will become idle during rescheduling.

Besides, there is a new trend to use deep reinforcement learning (DRL) techniques to learn dynamic scheduling policies \cite{mao2016resource,mao2019learning,gong2019chic}.
The DRL schedulers can automatically train the models based on offline traces, and dynamically fine-tune them in the online decision-making phase.
However, these DRL schedulers are not able to allow job preemption so far due to the limited action space size.
Therefore, it is questionable whether they are efficient to schedule DL jobs that are usually time-consuming.

\subsection{Advantages of the elasticity}
% \begin{figure}[t]
%     \centering
%     \begin{minipage}[t]{0.45\linewidth}
%         \centering
%         \captionsetup{width=0.9\linewidth}
%         \includegraphics[width=\linewidth]{figure/throughput.pdf}
%         \caption{Training speed of training ResNet50 on the CIFAR10 dataset.}
%         \label{fig:throughput}
%     \end{minipage}
%     \begin{minipage}[t]{0.45\linewidth}
%         % \centering
%         % \includegraphics[width=\linewidth]{figure/batchsize-exception.pdf}
%         % \caption{Scaling batch size from 256 to 4k at epoch 30 (training ResNet50 on the CIFAR10 dataset).}
%         % \label{fig:batch-exp}
%         \centering
%         \captionsetup{width=0.9\linewidth}
%         \includegraphics[width=\linewidth]{figure/fixed-local-bs.pdf}
%         \caption{Training accuracy with fixed local batch size of 256 and different number of GPUs.}
%         \label{fig:fixed-local-bs}
%     \end{minipage}
% \end{figure}
\begin{figure}[t]
    \centering
    \includegraphics[width=0.7\linewidth]{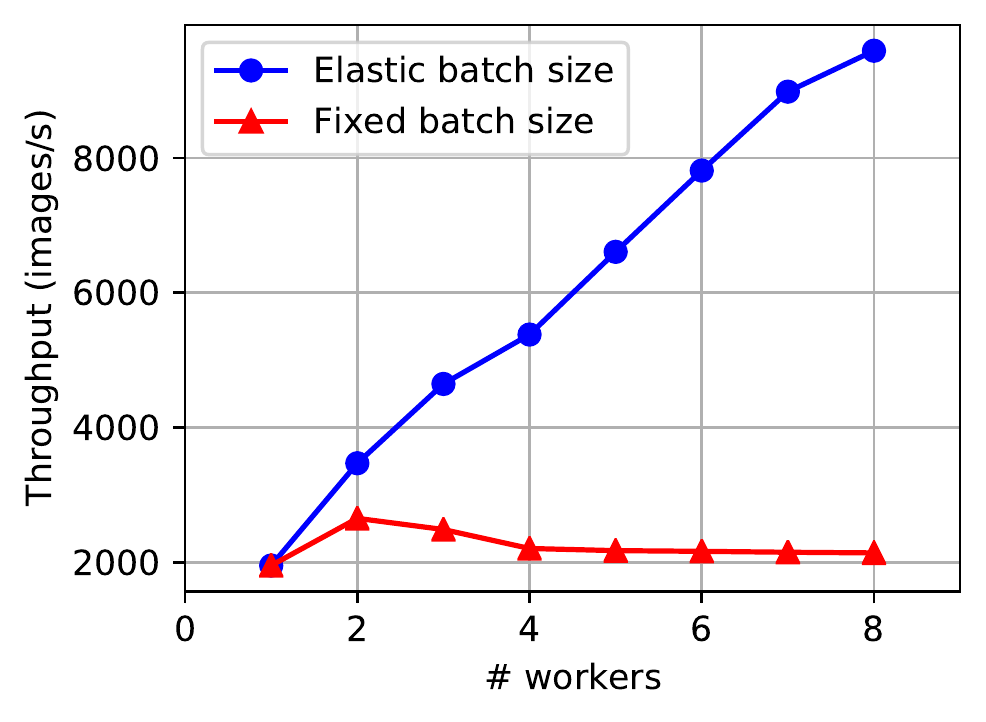}
    \caption{Training speed of training ResNet50 on the CIFAR10 dataset.}
    \label{fig:throughput}
\end{figure}
\begin{figure}[t]
    \centering
    \includegraphics[width=0.7\linewidth]{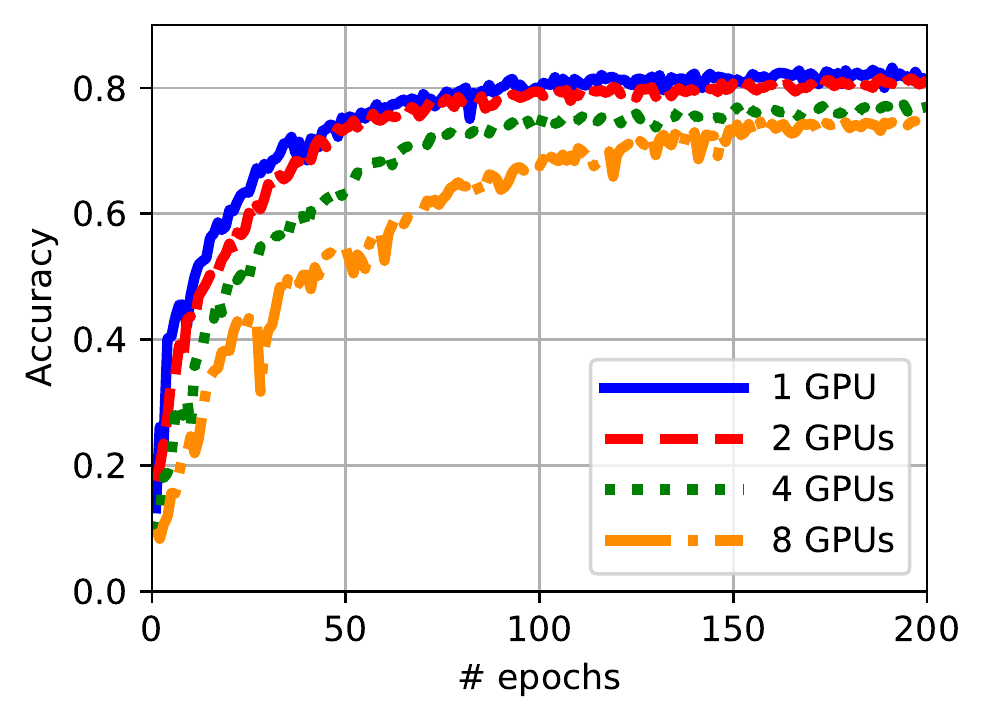}
    \caption{Training accuracy with fixed local batch size of 256 and different number of GPUs.}
    \label{fig:fixed-local-bs}
\end{figure}

The scheduler can make the best use of GPU resources to the advantage of the elastic batch sizes in different ways.
Through cautious batch size orchestration, DL jobs can find appropriate batch sizes to achieve higher GPU utilization, faster training speed and shorter job completion time.
For example, \autoref{fig:throughput} compares the speed of training ResNet50 \cite{he2016deep} on the CIFAR10 dataset \cite{cifar} between using an elastic batch size scaled from 256 to 2048 and a fixed batch size of 256.
However, with fixed global batch size, the utilization of each single worker will be reduced as the number of worker increases.
In the meantime, the throughput will be penalized by the increasing communication overhead.
As a result, when the number of workers exceeds 2, the throughput no longer increases and starts to drop.
Thus, to efficiently scale a job, the global batch size should increase as the number of workers increases, so that the throughput can keep increasing as well.
An existing common approach is to run workers with fixed local batch size on each GPU.
However, people use very large batch sizes with caution \cite{keskar2016large,hoffer2017train}.
With fixed local batch size, the global batch size will grow with the number of GPUs.
As shown in \autoref{fig:fixed-local-bs}, simply increasing the number of GPUs with a fixed local batch size can significantly affect the training convergence.
Using more GPUs leads to a slower convergence because the global batch size becomes larger, especially when the number of GPUs is greater than 2.
In order to guarantee training performance, selecting a proper learning rate will also be necessary when scaling the global batch size \cite{smith2017don,you2017large,you2019large}.

% \begin{figure}[t]
%     \centering
%     \includegraphics[width=0.6\linewidth]{figure/fixed-local-bs.pdf}
%     \caption{Training accuracy with fixed local batch size of 256 and different number of GPUs.}
%     \label{fig:fixed-local-bs}
% \end{figure}

Besides, using elastic batch sizes to train DL jobs can intuitively avoid fragmentation problem, where the number of idle GPUs is smaller than the required size of any pending job so that the GPUs are wasted.
In contrast, we allow the scheduler to execute some job with a smaller size first. As a result, we can saturate the cluster to fully utilize the GPU resources, and in the meantime reduce waiting time of the jobs.

\section{Online Evolutionary Scheduler}
\label{sec:design}
% To address the inefficiency in scheduling DL jobs, we aim to enhance their elasticity.
% We focus on synchronous distributed training with All-reduce communication that is the most popular combination of high performance distributed training,
% and design a scheduler that manages elastic batch size for each job and minimizes average job completion time with the optimal solution.

\begin{table}[t]
    \centering
    \begin{tabular}{c|c}
    \toprule
        Variable & Description\\
        \hline
        $J$ & the set of jobs \\
        \hline
        $C$ & the set of GPU devices\\
        \hline
        $c_j$ & the number of GPUs allocated to $j$\\
        \hline
        $B_j,b_j$ & the global \& local batch size of $j$\\
        \hline
        $T_j$ & remaining time of $j$\\
        \hline
        $Y_j$ & remaining workload of $j$\\
        \hline
        $X_j$ & throughput of $j$\\
        \hline
        $R_j$ & batch size limit of $j$\\
    \bottomrule
    \end{tabular}
    \caption{Summary of major variables.}
    \label{tab:note}
\end{table}

In this section, we introduce ONES for scheduling DL training jobs. 
The goal of ONES is to schedule the batch size of each job instead of the number of GPUs, and try to reach an appropriate balance between throughput and training performance by finding the optimal combination of batch size and number of GPUs.
More specifically, as shown in \autoref{tab:note}, let $J$ and $C$ denote the set of current jobs and GPU resources 
%%ww: (i.e., number of GPUs) 
in the cluster respectively, and let $c_j$ denote the number of GPUs allocated to each job $j$.
ONES maintains dynamic schedule decisions, which can be represented as the following mapping from jobs to resources:
\begin{gather}
    \label{eq:schedule} S:J \times C \to \{b_j^i\}, \quad b_j^i \in \mathbb{N}, ~\forall i \in C, ~\forall j \in J.
\end{gather}
where $b_j^i$ denotes the batch size for the worker of job $j$ on $\text{GPU}_i$.
We can also derive the global batch size and number of allocated GPUs as
\begin{gather}
    B_j = \sum_{i \in C} ~ b_j^i;\quad c_j = \sum_{i \in C} ~ \min (1, b_j^i).
\end{gather}
If a job has $B_j=c_j=0$, it cannot run and needs to wait for the next schedule.
Otherwise, the job will be running.

There are two core problems to be addressed:
(1) how to determine the optimal batch size for each job, and
(2) how to scale the batch size of a job with minimum cost.
We present our solutions in the rest of this section.

\subsection{Architecture}

% In this section, We present the design of ONES that is tailored for DL workloads to achieve the following goals:
% (1) improving elasticity so as to explore fine-grained scheduling decisions to reach the optimal performance,
% (2) minimizing the average job completion time of online workloads.
% Our designs focus on synchronous distributed training jobs with All-reduce communication, which serve as the dominant combination of high performance distributed training.

\begin{figure}[t]
    \centering
    \includegraphics[width=\linewidth]{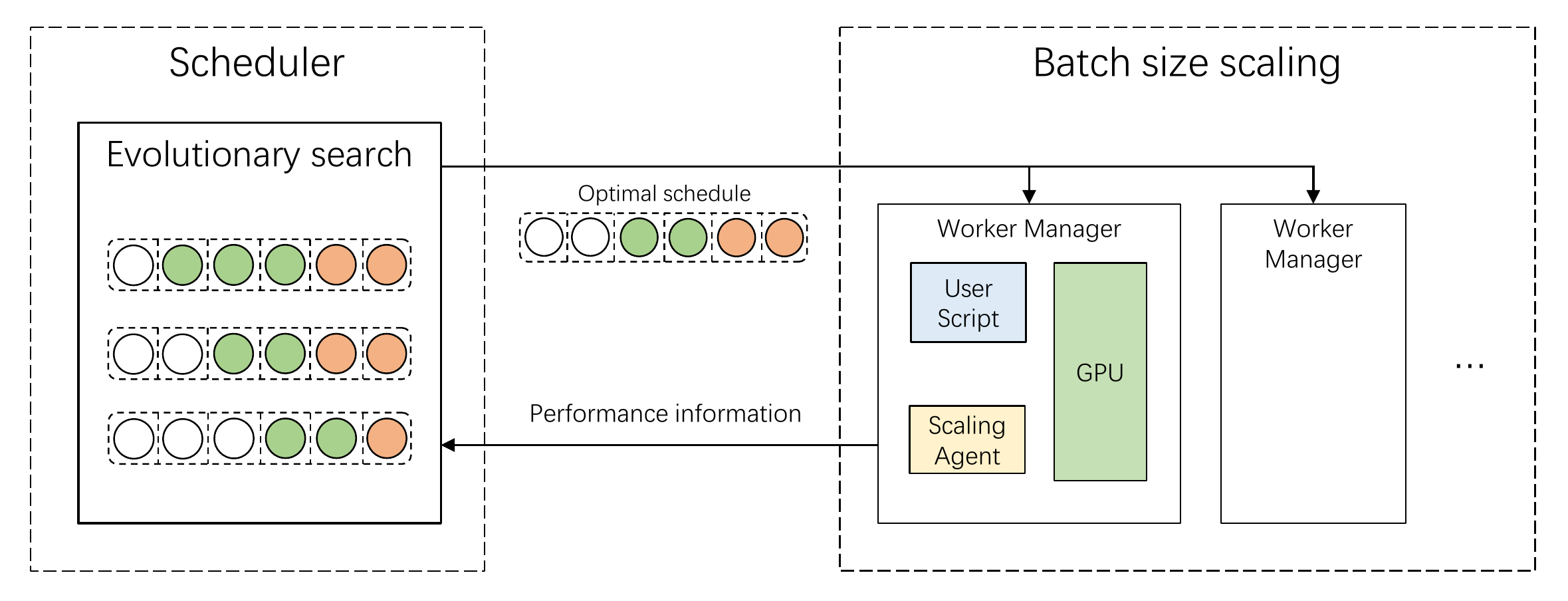}
    \caption{Overview of the scheduler architecture.}
    \label{fig:scheduler}
\end{figure}

ONES consists of two core functionalities as illustrated in \autoref{fig:scheduler}.
First, there is a central scheduler that dynamically \B{collects the real-time status of the cluster and jobs, and} determines the batch sizes of all the existing jobs.
The scheduler optimizes the solution based on the evolutionary search algorithm, in order to continuously reach high performance for online workloads.
% Then the scheduler selects a tentatively optimal solution from time to time, and executes it on the cluster.
\B{Then the scheduler allocates available GPUs based on the solution, and invokes corresponding workers to execute the decision on the GPU nodes.}

Next, to execute the new solution, we may need to re-configure some jobs with new batch sizes.
However, the common practice is to stop the training and restart them with new configurations, which can result in significant overheads.
We present a new approach to execute batch size scaling that is almost invisible to the original training.
\B{More specifically, a worker manager is bound to each GPU device, which} receives the new configuration from the scheduler, and invokes a scaling agent to automatically adjust the execution configurations of its worker in the background.
Thus, the batch size scaling no longer needs to stop the training.

Besides, each worker uploads its training progress \B{(e.g. number of processed samples, training loss and validation accuracy)} to the central scheduler at the end of each training epoch.
\B{Each job ends when its model converges.
Then the scheduler will release its workers and clear its resource allocation.}

\begin{figure}[t]
    \centering
    \includegraphics[width=0.75\linewidth]{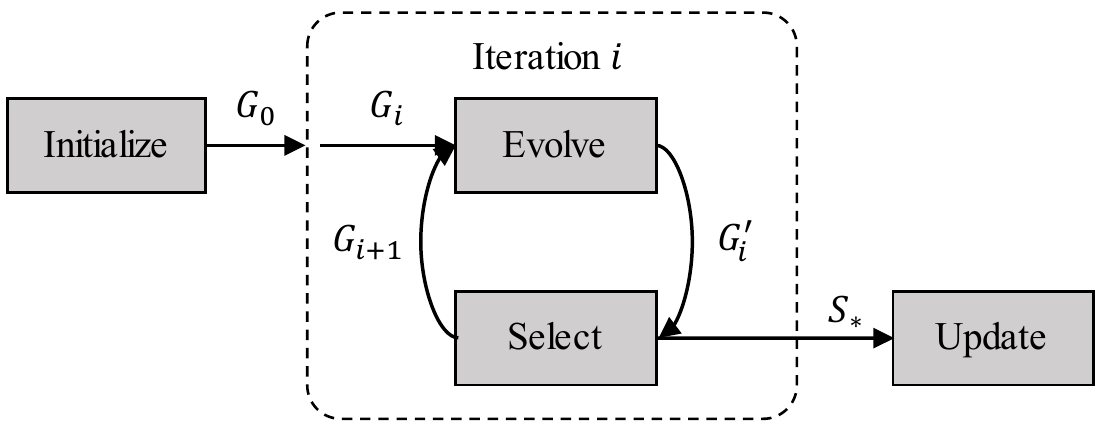}
    \caption{The online evolutionary scheduling algorithm. $G_0$: the initial population; $G_i$ the current population; $G'_i$: new solutions derived from $G_i$; $G_{i+1}$: solutions with higher scores selected from $G'_i$; $S_*$: the best solution.}
    \label{fig:evolution}
\end{figure}

\subsection{Online Evolutionary Search}
\label{sec:online}

To determine appropriate batch sizes for the DL jobs that continuously arrive and compete for a limited amount of GPU resources, we design an online scheduler based on evolutionary search to minimize job completion time (JCT).

We present the workflow of the algorithm as illustrated in \autoref{fig:evolution}.
The algorithm optimizes a population that consists of a group of candidate solutions with an iterative evolution process.
It starts with an initial population $G_0$.
%%ww: give an example or definition of "population".
In each iteration $i$, new candidates are derived from the current population $G_i$, and those with higher scores will be selected to form the new population $G_{i+1}$.
Finally, the schedule will be updated by deploying the optimal candidate $S_*$.

\B{The evolutionary search can be more suitable than existing scheduling strategies that make greedy decisions to schedule DL workloads in the following aspects.
First, the algorithm continuously optimizes a population of solutions that reduce the chance to be trapped in local minima.
Thus it can achieve higher-quality solutions than greedy strategies that optimize a single solution.
Second, the population of solutions keeps evolving according to real-time job status, which helps ONES make immediately response to online workloads, especially when GPU resources become available after job completion, or very short jobs arrive and provisionally taking over some GPUs from long jobs can be beneficial.
Besides, compared with other approximate search algorithm such as Simulated Annealing (AS), Tabu Search (TS), Nearest Neighbor Search (NNS) and Ant Colony Optimization (ACO), the evolutionary search is the most intuitively suitable for the scheduling problem as the allocation of GPUs can be directly encode into genomes (as illustrated in \autoref{fig:encoding}). In contrast, it is not very intuitive to define the “neighborhood” for NNS or the “path” for ACO.
In the meantime, the evolutionary search has relatively fast iterative speed.}
% Unlike greedy approaches, it works with a population of solutions instead of a single one that may usually be trapped in a local optimum.
% Besides, the algorithm will be efficient in dealing with online workload by continuously searching for more fit solutions to make quick responses to any change in the cluster.

\subsubsection{What Candidate Wins?}\hfill\\\indent
\label{sec:sruf}
% The principle of the evolutionary search is continuously selecting the solutions with higher performance.
% Thus we first need to define how to evaluate the performance of each candidate solution.
% The evaluation aims to select solutions that can induce smaller average JCT.
The goal of ONES scheduling algorithm is to minimize the average JCT.
Preferentially, serving the job with the \textit{shortest remaining processing time} (SRPT) is the solution to this problem. \cite{crovella1999connection,tetris,grosof2018srpt}.
Sadly, this approach may lead to a potential consequence that the scheduler allocates as many GPUs as possible to each job.
This is apparently not sensible because the training speed of distributed DL jobs does not scale linearly with the number of GPUs due to a considerable communication overhead in parameter synchronization.
Therefore, we extend SRPT and define \textit{smallest remaining utilization first} (SRUF) as our goal.
Let $J_r$ denote the running jobs, and $T_j$ denote the remaining processing time of job $j \in J_r$. Note that $T_j$ is supposed to be a function of the batch size $B_j$. 
We can represent the utilization of job $j$ w.r.t both time and space as $T_j(B_j) \times c_j$.
Then, we solve the optimization problem to determine the batch size for each job by minimizing the overall utilization as follows:
\begin{align}
    \label{eq:sruf}            \text{minimize} &~ \sum_{j \in J_r} ~ T_j(B_j) \times c_j, \\
    % \label{eq:sruf_constraint} \text{subject to} &~ \sum_{j \in J_r} ~ c_j = \|C\|, \\
    \label{eq:surf_exclusive}  \text{subject to} &~ \sum_{j \in J_r} ~ \min(1,b_j^i) = 1, \forall i \in C.
\end{align}
% Note that we only consider the solutions where all the GPUs are allocated (\autoref{eq:sruf_constraint}), because with the elastic batch size, we can avoid fragmentation problem and maximize GPU utilization.
Note that we do not allow multiple jobs to share a single GPU device due to the severe interference caused by GPU sharing \cite{jeon2019analysis}, i.e. each worker of the jobs will be exclusively executed on a single GPU (\autoref{eq:surf_exclusive}).

However, it is difficult to obtain the remaining time of each job because we can hardly know how many samples to train until a DL job is completed.
Let $X$ denote the throughput (i.e. training speed) and $Y$ denotes the remaining workload (\# samples to process until convergence), we compute the remaining time as
\begin{align}
    \label{eq:remaining_time} T= \frac{Y}{X},
\end{align}
We can easily profile real-time throughput by measuring the throughput on each GPU during training, and use the mean value of collected measures in computation.
To obtain $Y$, we take advantage of the cluster history and try to excavate useful information.
The key insight is that, despite the difficulty to know the exact remaining workload size, we may be able to model probability distribution of it based on historical jobs that have similar training performance.
% Then with the probability distributions, we can select the optimal solution with minimum expected remaining utilization by probability sampling.

\begin{figure}[t]
    \centering
    \includegraphics[width=0.8\linewidth]{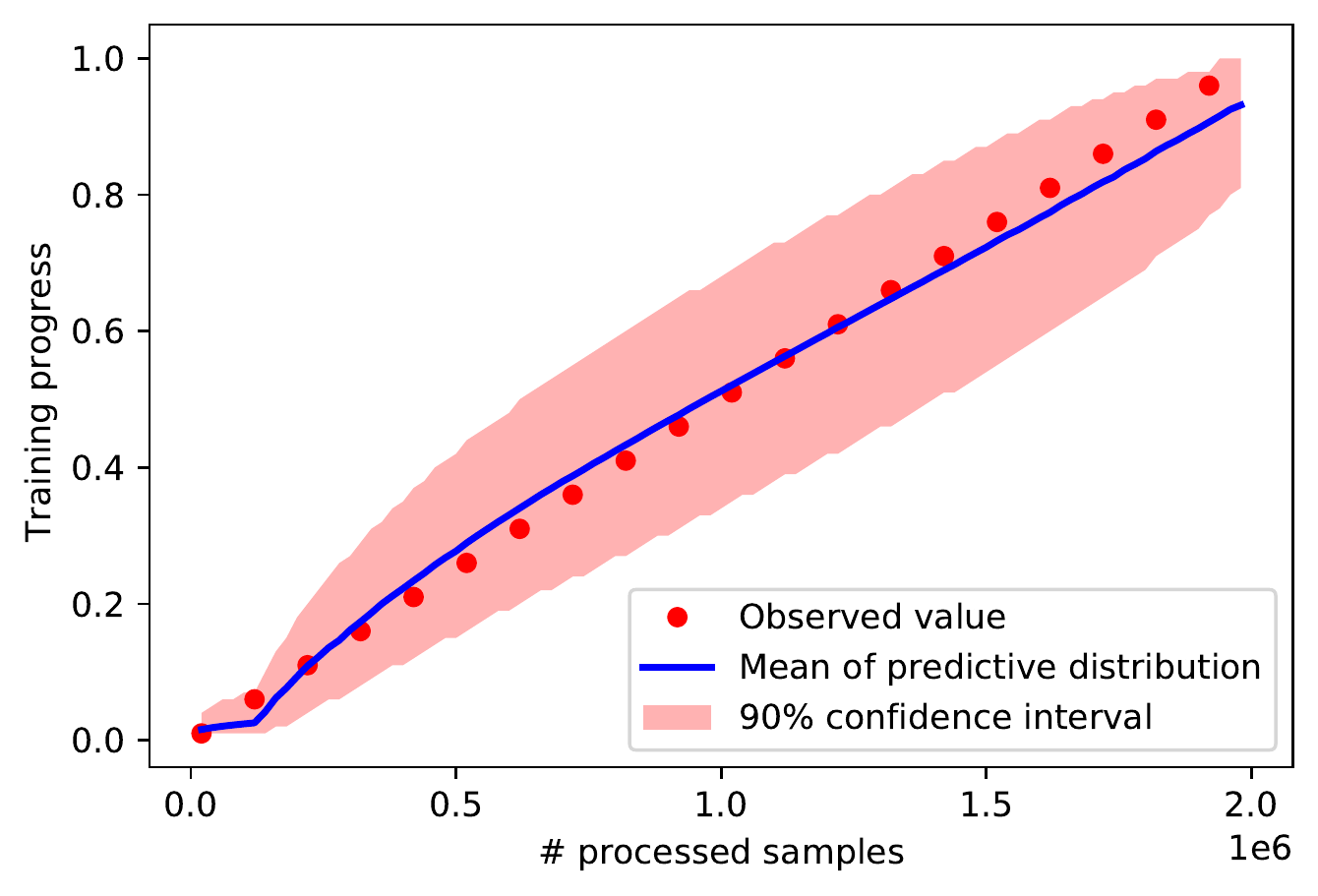}
    \caption{An example of online prediction progress with uncertainty.}
    \label{fig:predict}
\end{figure}

To obtain the distributions for online workloads, we conduct the following online prediction process.
The exact numbers of workload of DL jobs have a wide range, so that it is not appropriate to directly predict the workload.
Instead, we predict the training progress (i.e. percentage of completion) that can be limited in $(0,1)$.
Then, we use Beta distributions to model the uncertainty, because Beta distributions can produce flexible shapes between 0 and 1.
Moreover, the shape of a Beta distribution $Be(\alpha, \beta)$ is unimodal when $\alpha,\beta>1$, which fits the law of large numbers.
Let $\rho$ denote the progress, we can build the following regression model to predict its distribution.
For any job with input features $x$
\footnote{
We use
$x=\{\|D\|, \mathcal{L}_{initial}, Y_{processed}, r_{\mathcal{L}}, \mathcal{A}\}$
that can be captured from the training history of completed jobs as input features of the GPR predictor,
where $\|D\|, \mathcal{L}_{initial}$ represent the epoch size and initial loss (before training) of the job respectively, $Y_{processed}$ represents the number of already processed samples, $r_{\mathcal{L}}$ represents the loss improvement ratio, computed as
$r_{loss} = 1-\text{current loss}/\text{initial loss}$, and $\mathcal{A}$ represents the validation accuracy.
}
, we assume its training progress is an independent random variable that
\begin{gather}
    \rho \sim Be(\alpha, \beta) \\
    \nonumber
    \begin{aligned}
        \text{where} ~& \alpha=Y_{processed}/\|D\|, \\
                    &\beta=\max(\textbf{A} x+b, 1),
    \end{aligned}
\end{gather}
which approximate the number of processed epochs and epochs to process respectively, defined by a pair of parameters $A$ and $b$.
We apply a threshold function to both $\alpha$ and $\beta$ to guarantee $\alpha, \beta \ge 1$.
%%ww: are you fitting the progress (i.e., loss) curve? what are the inputs and outputs?

ONES continuously updates the regression model for online prediction.
Each time when a job is completed, we train the model by maximizing the log marginal likelihood to fit the data collected from historical job information.
Note that we maintain a limited size of training dataset where the data points are uniformly sampled from training logs of completed jobs.
By doing so, we can control a reasonable training time and prevent overfitting.

\begin{algorithm}[t]
    \caption{Probability sampling}
    \label{alg:sampling}
    \LinesNumbered
    \SetKwInOut{Input}{Input}
    \SetKwInOut{Output}{Output}
    \SetKw{Break}{break}
    \Input{$S_1, ..., S_K;Be_1, ..., Be_J$}
    \Output{The selected sample $S_*$}

    \For{$j=1$ \KwTo $J$}{
        sample $\rho_j$ from $Be_j$ \label{line:sample}
    }
    \For{$i=1$ \KwTo $K$}{
        $
        \begin{aligned}
            score_i \gets \sum_{j \in J_r} ~ \frac{{Y_{processed}}_j c_j}{X_j} (\frac{1}{\rho_j}-1)
        \end{aligned}
        $ \label{line:score}
    }
    $S_* \gets S_i$ if $score_i \le score_k$, $\forall k\in K$ \label{line:select}
    
\end{algorithm}

Then, the regression model is used to predict the distribution of $\rho_j \sim Be_j$, for online workloads $\forall j \in J$.
\autoref{fig:predict} shows an example of the prediction where the line illustrates the mean values of the output distributions, while the shadowed area represents the 90\% confidence intervals of the distributions.
Next, we can derive the remaining workload $Y_j$ as
\begin{align}
    \label{eq:y} Y_j = {Y_{processed}}_j (\frac{1}{\rho}-1),
\end{align}
where $ {Y_{processed}}_j$ denotes the number of samples that are already processed by job $j$. 
%%ww: in job j?
Applying \autoref{eq:y} to \autoref{eq:remaining_time}, we can derive the overall utilization of each solution as
\begin{align}
    \label{eq:predicted_util} \sum_{j \in J_r} ~ T_j \times c_j = \sum_{j \in J_r} ~ \frac{{Y_{processed}}_j c_j}{X_j} (\frac{1}{\rho_j}-1)
\end{align}

Finally, we can select an expected optimal solution as illustrated in \autoref{alg:sampling}.
The sampling algorithm draws samples
%%ww: may need to explain the algorithm a bit. What is a sample?
from the training progress distributions of each job (\autoref{line:sample}) to compute the score of each solution (\autoref{line:score}).
Then the solution with the smallest score is selected (\autoref{line:select}).

\subsubsection{Search Process}\hfill\\\indent
We present the details of the evolutionary search process in \autoref{fig:evolution} as follows.

\textbf{Initialization:}
We generate an initial population $G_0$ with $K$ simple solutions.
We suggest to use the same size for the population as the cluster.
Thus, the initial solutions can be easily generated by running a random job on each GPU respectively.

\textbf{Evolution:}
The algorithm then generates new candidate solutions by randomly adjusting the batch sizes of the jobs.
We design four random operations to evolve candidates from the current population $G_i$, including the \textit{refresh}, \textit{uniform crossover}, \textit{uniform mutation} and \textit{reorder} operations.
% Since scaling the batch sizes of DL jobs without restraint may lead to convergence issues \cite{keskar2016large,hoffer2017train}, 
Note that we set a limit to the batch size as $R_j$ of each job $j \in J$ to avoid convergence issue.
$R_j$ is dynamically updated according to the real-time training performance.
The details will be described in \S \ref{sec:policy}.
% As a result, the evolution is guaranteed to be online because the decisions are made based on the information such as the batch size limit $R$ that is updated at real time.

The evolution operations are executed as follows.

% \begin{algorithm}[!t]
%     \caption{Evolution}
%     \label{alg:evolve}
%     \LinesNumbered
%     \SetKwInOut{Input}{Input}
%     \SetKwInOut{Output}{Output}
%     \Input{The current population $G_i$,
%           jobs $J$, batch size limits $R$,
%           population size $K$}
%     \Output{The new candidates $G_i'$}
%     \SetKwFunction{Refresh}{refresh}
%     \SetKwFunction{Crossover}{crossover}
%     \SetKwFunction{Mutate}{mutate}
%     \SetKwFunction{Reorder}{reorder}
%     \SetKwProg{Fn}{function}{}{end}
    
%     \tcp*[h]{Set up} \\
%     \For{$S \in G_i$}{
%         $S \gets$ \Refresh($S, J, R$) \\
%         Save $S$ to $G_i'$
%     }
%     \tcp*[h]{Crossover} \\
%     \For{$j=1$ \KwTo $\frac{K}{2}$}{
%         Randomly select $S_a$ and $S_b$ from $G_i$ \\
%         $S_c, S_d \gets $ \Crossover($S_a, S_b, J, R$) \\
%         Save $S_c$ and $S_d$ to $G_i'$
%     }
%     \tcp*[h]{Mutation} \\
%     \For{$S \in G_i$}{
%         $S' \gets$ \Mutate($S, J, R$) \\
%         Save $S'$ to $G_i'$
%     }
%     \tcp*[h]{Reorder} \\
%     \For{$S' \in G_i'$}{
%         \Reorder($S'$)
%     }
% \end{algorithm}

\begin{figure}[!t]
    \centering
    \includegraphics[width=0.6\linewidth]{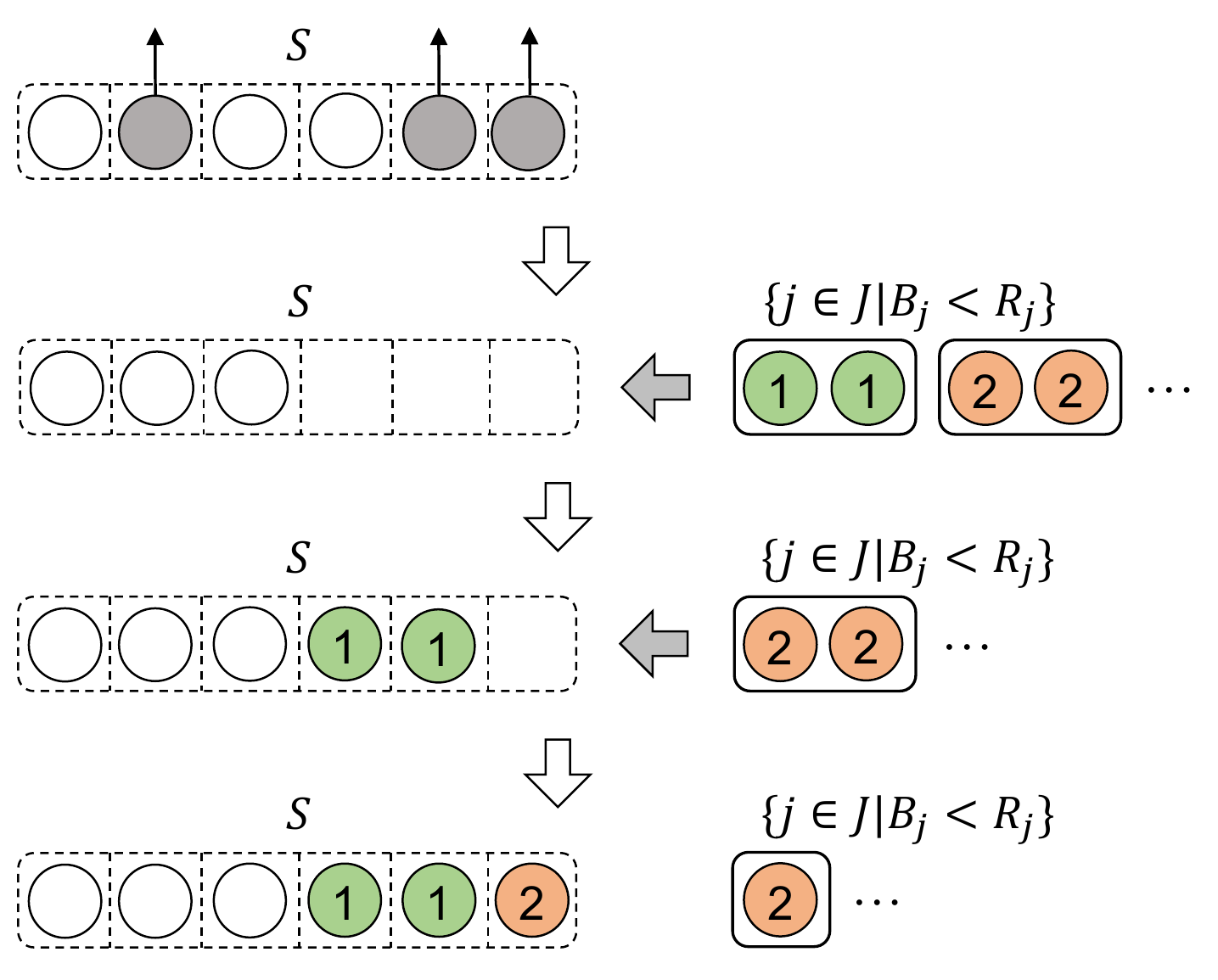}
    \caption{An example of refreshing the cluster and filling the idle GPUs. Job 1 is firstly selected by probability sampling and there is still an idle GPU left; then Job 2 is selected so that the cluster is full and the \textit{refresh} operation end.}
    \label{fig:insert}
\end{figure}

\begin{figure}[!t]
    \centering
    \includegraphics[width=0.75\linewidth]{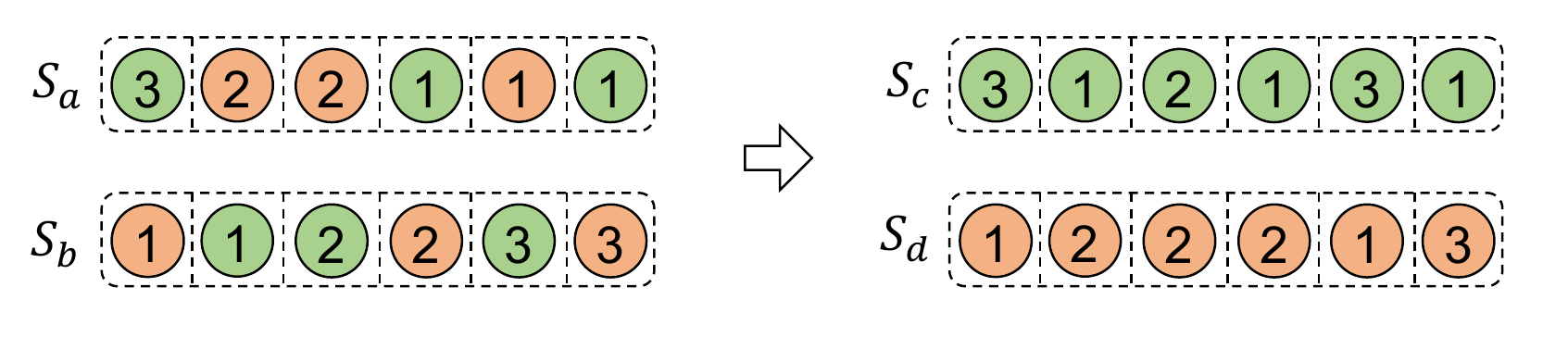}
    \caption{An example of the uniform crossover operation. The left side: parent candidates; the right side: child candidates.}
    \label{fig:crossover}
\end{figure}

\begin{figure}[!t]
    \centering
    \includegraphics[width=0.5\linewidth]{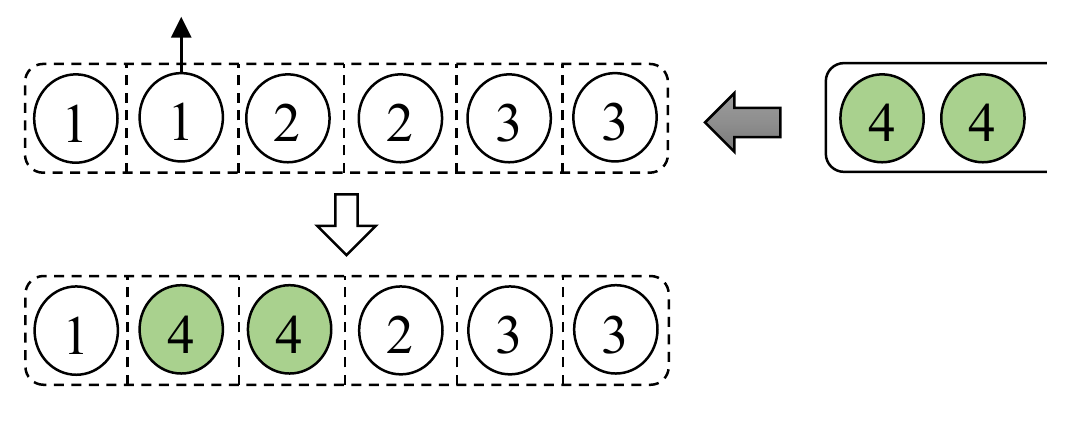}
    \caption{An example of the uniform mutation operation. Two original workers are replaced by Job 4.}
    \label{fig:mutate}
\end{figure}

\begin{figure}[!t]
    \centering
    \includegraphics[width=0.7\linewidth]{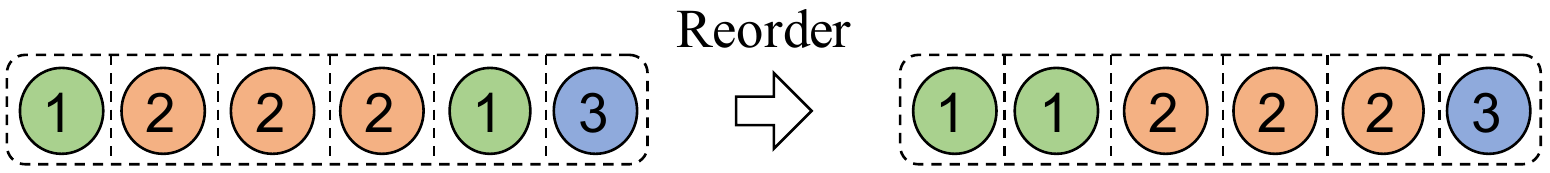}
    \caption{An example of the reorder operation. The left side: before reorder; the right side: after reorder.}
    \label{fig:reorder}
\end{figure}

\begin{itemize}
    \item \textit{Refresh}:
    This operation is designed to update the schedules in $G_i$ with the real-time job status.
    Thus, in this step, ONES
    (1) cleans up the GPU resources where the jobs have been completed
    (2) scales down any job $j \in J_r$ by $c_j - \lfloor \frac{R_j \cdot c_j}{B_j}\rfloor$ GPUs if $R_j < B_j$, and
    (3) allocates $N$ new jobs with $N$ available GPUs.
    
    When the number of available GPUs in a schedule is smaller than $N$ after (1) and (2), the scheduler will take GPUs from the jobs with the largest $T_{processed}$ until $N$ new jobs are fully allocated.
    We can thereby avoid starvation of new jobs by the preferential allocation.
    
    Furthermore, when there are still idle GPUs after (1) (2) and (3), we will need to fill them by resuming waiting jobs or scaling up running jobs.
    We compare the jobs by computing the utilization gain of each job $j \in J$ after increasing its batch size from $B_j$ to $R_j$ with $\lfloor \frac{R_j \cdot c_j}{B_j}\rfloor - c_j$ more GPUs.
    Then, as illustrated in \autoref{fig:insert}, we can continuously select one job by probability sampling from $\{\Delta \varphi_j Y_j\}$ using \autoref{alg:sampling} until there is no idle GPU.
    
    \item \textit{Uniform crossover}:
    This operation is designed to combine two parent candidates to generate two child candidates.
    More specifically, it scans over each GPU $i \in C$ identically and independently.
    On each GPU, the job from the first parent is inherited by a random child, and the job from the second parent is inherited by the other child, as shown in \autoref{fig:crossover}.
    % Note that inheriting a job $j$ cannot exceed its batch size limit $R_j$.
    % \autoref{fig:crossover} shows an example of crossing over $S_a$ and $S_b$ to generate $S_C$ and $S_d$, where the jobs with workers in the green circles are inherited by $S_c$, and those in the orange circles are inherited by $S_d$.
    In each iteration, $K$ pairs of parents are randomly selected from the original population, and $K$ pairs of children will be generated.
    
    \item \textit{Uniform mutation}:
    This operation is designed to exchange the GPU resources of some jobs in a candidate schedule with other jobs.
    More specifically, $K$ candidates will be randomly selected from the original population, where each job in each candidate $S$ will be randomly preempted with a mutation rate $\theta$.
    The GPU resources of these preempted jobs will be filled with waiting jobs or other running jobs selected as shown in \autoref{fig:mutate}.
    
    \item \textit{Reorder}:
    The operations above may generate candidates with poor placement because workers of the same job are not guaranteed to be closely located.
    Therefore, we design the \textit{reorder} operation to pack the workers of the same job in the order of the occurrence of each job, as illustrated in \autoref{fig:reorder}.
    
\end{itemize}

\textbf{Selection:}
In the set of new candidate schedules $G_i'$ generated by the evolution operations, the score of each candidate is derived using Eq. \ref{eq:predicted_util}, based on the predicted distribution of training progress $Be_1, ..., Be_J$.
Then the best $K$ candidates are selected from $G_i'$ to form the new population $G_{i+1}$ by probability sampling that is similar to Alg. \ref{alg:sampling} but will draw $K$ samples with the top scores.

\textbf{Update:}
The solution with the highest score $S_*$ will be executed in the cluster.
However, too frequent update may reduce the scheduling performance to some degree due to the overhead of adjusting resource allocation and job placement.
At least, we have to ensure that each job can finish one epoch so as to completely scan its training dataset.
Therefore, ONES updates the schedule with the optimal candidate $S_*$ after all the running jobs finish at least one epoch.

\subsection{Elastic Batch Size Scaling}
\label{sec:scaling}

To execute the scheduling solutions with minimum cost and without affect on the training performance, we present elastic batch size scaling that is based on the analysis of common DL training practices.

\begin{figure*}[t]
    \begin{subfigure}[t]{0.27\textwidth}
        \centering
        \includegraphics[width=\linewidth]{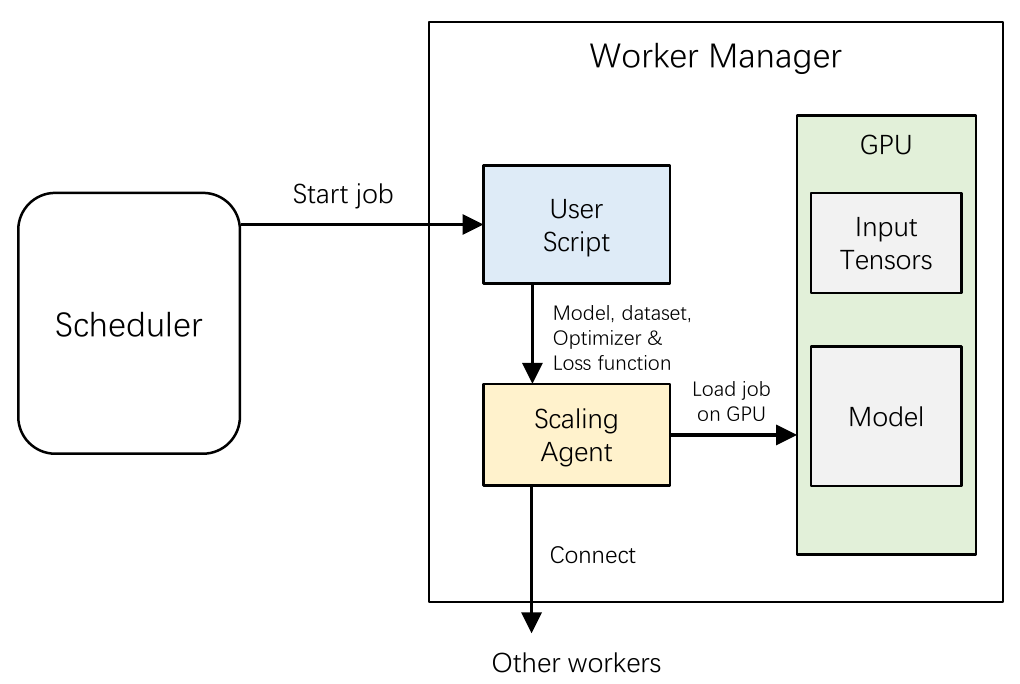}
        \caption{}
        \label{fig:orche-a}
    \end{subfigure}
    \begin{subfigure}[t]{0.2\textwidth}
        \centering
        \includegraphics[width=0.8\linewidth]{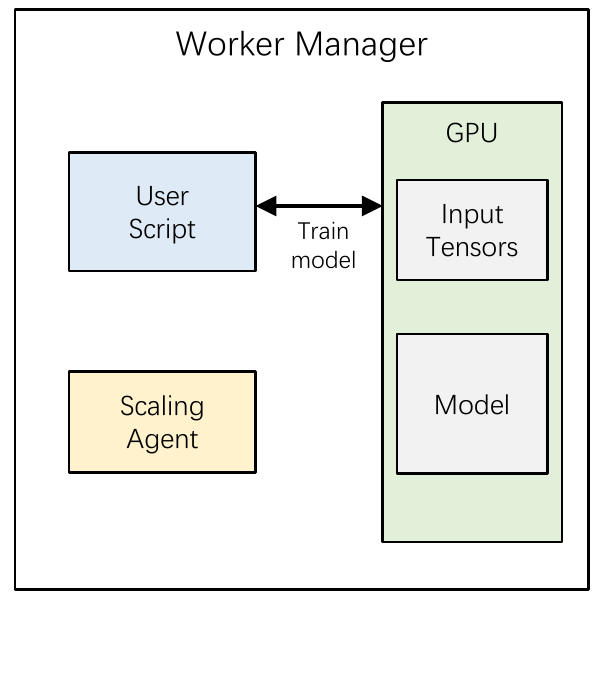}
        \caption{}
        \label{fig:orche-b}
    \end{subfigure}
    \begin{subfigure}[t]{0.27\textwidth}
        \centering
        \includegraphics[width=\linewidth]{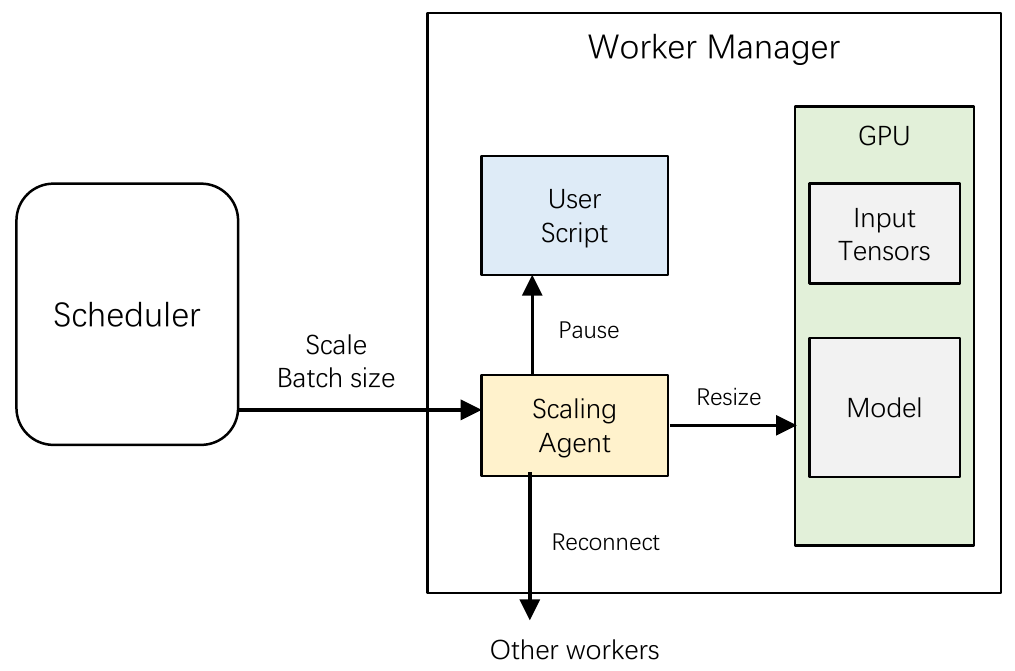}
        \caption{}
        \label{fig:orche-c}
    \end{subfigure}
    \begin{subfigure}[t]{0.2\textwidth}
        \centering
        \includegraphics[width=0.8\linewidth]{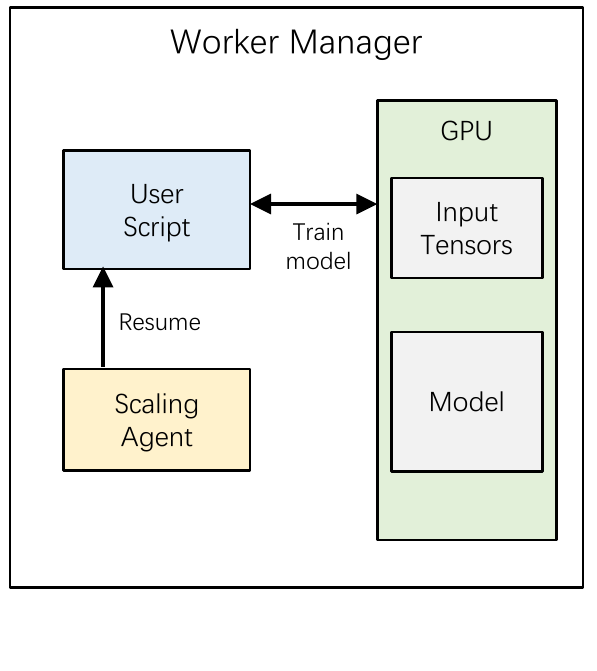}
        \caption{}
        \label{fig:orche-d}
    \end{subfigure}
    \caption{Scaling the batch size in 4 steps.}
    \label{fig:orchestration}
\end{figure*}

\begin{figure}[t]
    \centering
    \includegraphics[width=0.9\linewidth]{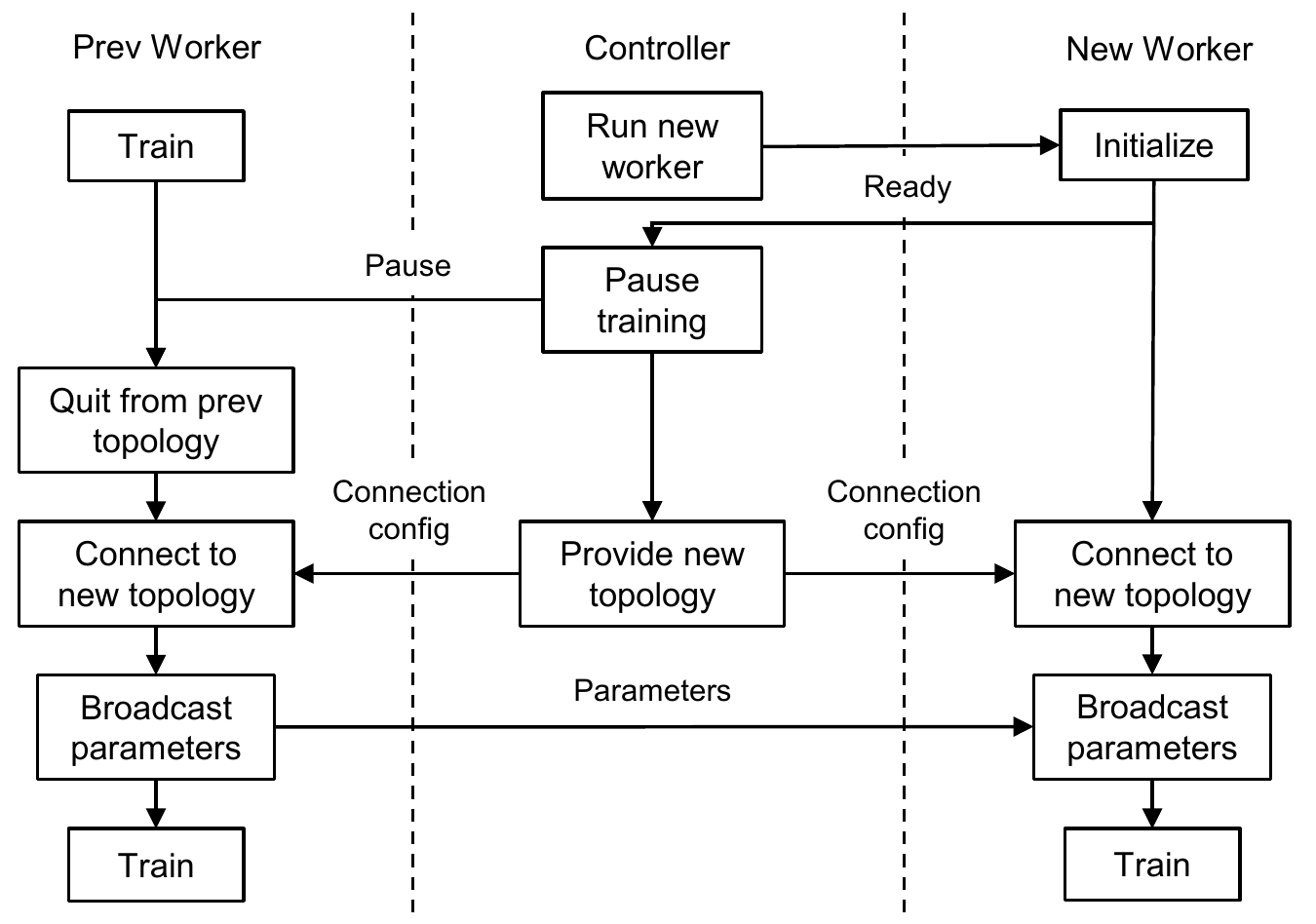}
    \caption{Workflow of adding a new worker.}
    \label{fig:ckpt-free}
\end{figure}

\subsubsection{Batch Size Scaling Mechanism}\hfill\\\indent
\label{sec:orchest}
The batch size scaling re-configures a job with a new batch size.
The re-configuration is generally not performed in a direct way, because the change in batch size involves not only itself but also the distributed model replicas, the input tensors and the input data batches.
Today the dominant practice is to save and stop the training, and then restart it with the new configuration, which usually takes tens of seconds to complete.
Gu \emph{et al} \cite{gu2019tiresias} provide the detailed numbers of the overhead of migrating Tensorflow \cite{abadi2016tensorflow} jobs with checkpoints, which are basically around tens of seconds.
Our scaling mechanism avoids stopping the training and completes the scaling in background in the following four steps, as illustrated in \autoref{fig:orchestration}.

Initially, the scheduler binds several worker managers to a job, and start the job on them. Each worker manager starts the user script to initialize the job, and invokes a scaling agent to load the modules to execute on the GPU device.
The scaling agent also connects with other workers of the same job for distributed communication.

Secondly, after the initialization, the user script is allowed to start the training.

Next, when the job needs scaling, the scheduler inform its worker managers of the new configuration.
Then the scaling agent pauses the user script at the end of a training step, resizes the modules in the GPU devices, and reconnects other workers.
If there are new workers added to the job, we will need to share the current model parameters with them.
To avoid unnecessary waiting time, we can start the new workers first and overlap their initialization with previous training.
As illustrated in \autoref{fig:ckpt-free}, the new workers first start with initialization.
When they are ready, the workers will notify all the previous workers via the controller.
After the notified previous workers complete a training step, they quit from the previous topology.
Then all the workers connect to the new topology together, and broadcast the current parameters together from one of the previous workers.

Finally, after the scaling, the scaling agent resumes the user script to continue the training with the new configuration.

\subsubsection{Training Performance Control}\hfill\\\indent
\label{sec:policy}
To implement elastic batch size scaling, we still need to address several underlying performance problems so as not to affect the original training convergence.

Empirically, the batch size is carefully selected to reach high accuracy, so that adjusting the batch size may affect the training performance.
According to the analysis of prior studies \cite{keskar2016large,hoffer2017train}, with a very large batch size, a job will take more epochs to converge and even reach a worse accuracy.
However, some recent works \cite{smith2017don,goyal2017accurate} observe that by carefully adjusting the learning rate, DL jobs that use gradient descent optimizers can still reach equivalent accuracy after the same number of training epochs.
They suggest to linearly scale the learning rate with the batch size, because for example by increasing the batch size from $B$ to $kB$, the number of steps will be reduced by $k$, so that the number of updates will be reduced by $k$; in the meantime the learning rate is increased from $\eta$ to $k\eta$, which will also enlarge each update by $k$. Therefore, the training can still achieve an equivalent progress.

% \begin{figure}[t]
%     \centering
%     \begin{minipage}[t]{0.45\linewidth}
%         \centering
%         \captionsetup{width=0.9\linewidth}
%         \includegraphics[width=\linewidth]{figure/batchsize-exception.pdf}
%         \caption{Scaling batch size from 256 to 4k at epoch 30 (training ResNet50 on the CIFAR10 dataset).}
%         \label{fig:batch-exp}
%     \end{minipage}
%     \begin{minipage}[t]{0.45\linewidth}
%         \centering
%         \captionsetup{width=\linewidth}
%         \includegraphics[width=\linewidth]{figure/batchsize-loss.pdf}
%         \caption{Training loss of scaling batch size from 256 to 4096 gradually. $B=256$ in the first 30 epochs; $B=1024$ in the next 30 epochs; $B=4096$ in the last 30 epochs.}
%         \label{fig:scaling}
%     \end{minipage}
% \end{figure}

\begin{figure}[t]
    \centering
    \includegraphics[width=0.7\linewidth]{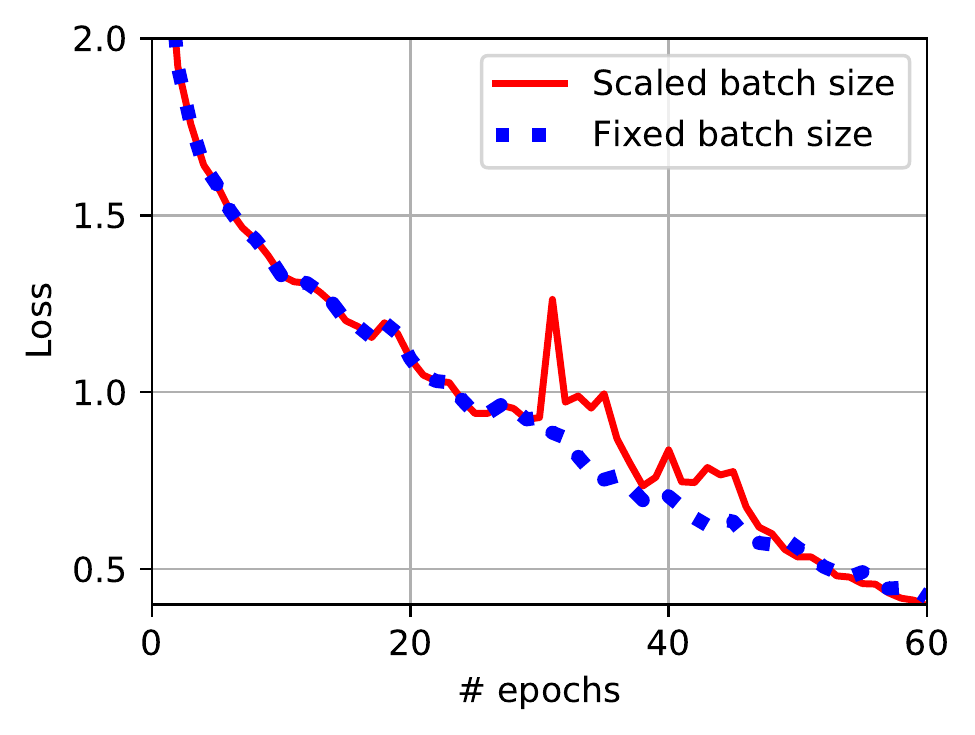}
    \caption{Scaling batch size from 256 to 4k at epoch 30 (training ResNet50 on the CIFAR10 dataset).}
    \label{fig:batch-exp}
\end{figure}

\begin{figure}[t]
    \centering
    \includegraphics[width=0.7\linewidth]{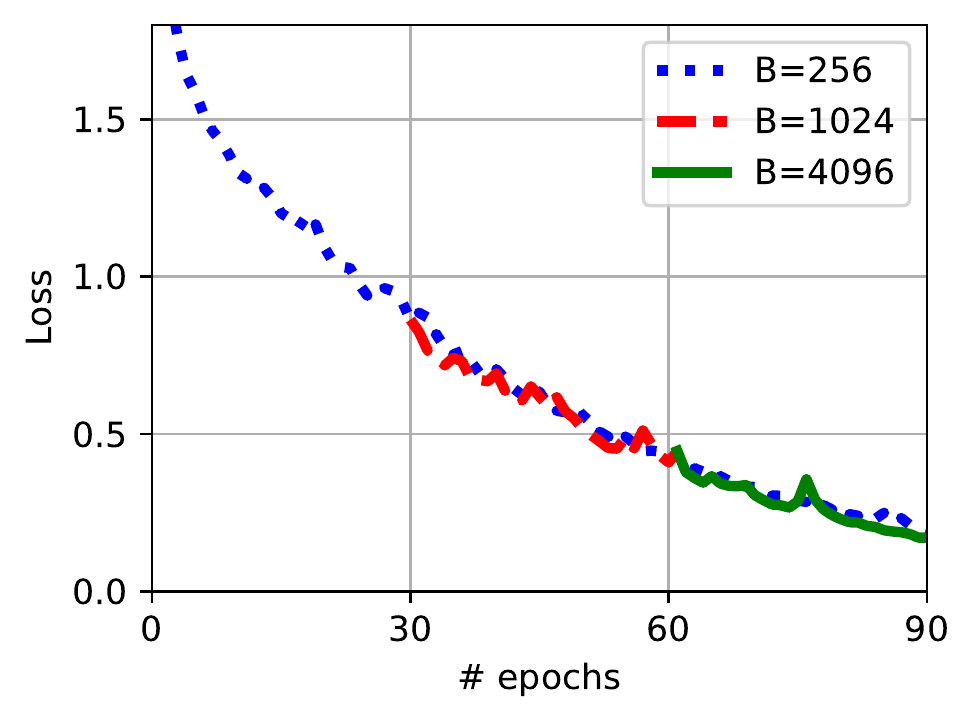}
    \caption{Training loss of scaling batch size from 256 to 4096 gradually. $B=256$ in the first 30 epochs; $B=1024$ in the next 30 epochs; $B=4096$ in the last 30 epochs.}
    \label{fig:scaling}
\end{figure}

Another problem that affects the training performance occurs when there is a too large gap of batch size before and after scaling.
As shown in \autoref{fig:batch-exp}, such scaling results in a sudden increase in the training loss.
This is caused by the noise in gradient and momentum states when the batch size explodes \cite{lin2019dynamic}.
After the increase, the training will need many epochs to return the equivalent level of normal performance.
In contrast, if the batch size grows gradually into a larger value, as shown in \autoref{fig:scaling}, the training performance can be basically guaranteed.
Therefore, we only allow the batch size to be scaled within a limited range at each time.

% \begin{figure}[t]
%     \centering
%     \begin{subfigure}[t]{0.48\linewidth}
%         \centering
%         \includegraphics[width=\linewidth]{figure/batchsize-loss.pdf}
%         \caption{Training loss}
%     \end{subfigure}
%     \begin{subfigure}[t]{0.48\linewidth}
%         \centering
%         \includegraphics[width=\linewidth]{figure/batchsize-acc.pdf}
%         \caption{Validation accuracy}
%     \end{subfigure}
%     \caption{Training ResNet50 \cite{he2016deep} on the CIFAR10 dataset \cite{cifar} with batch size scaled from 256 to 4096. $B=256$ in the first 30 epochs; $B=1024$ in the next 30 epochs; $B=4096$ in the last 30 epochs.}
%     \label{fig:scaling}
% \end{figure}

In general, ONES jointly manages the batch size and learning rate of each job according to their initial values based on linear scaling, so that an equivalent training performance can be presented as expected.

To control the training performance, we set a dynamic limit to the batch size of each job, denoted as $R$ so that the scheduler will not allow the job to exceed the limit.
As the training proceeds, $R$ will be adjusted according to the following scaling policies that depend on different situations.
\begin{itemize}
    \item \textit{Start:}
    Upon arrival, a job will have to limit its batch size to be accommodated in a single GPU until completing a few warm-up steps.
    \item \textit{Resume:}
    A job that is waiting for service can request the batch size limit $R$ that does not exceed the amount before being preempted, but once it is rejected (i.e. the job will still be waiting in the next schedule), $R$ will be halved.
    The purpose is to reduce the queuing time and prevent starvation.
    %%ww: what is R? not very clear.
    % A: R is the maximum batch size to scale, which is mentioned above
    \item \textit{Scale-up:}
    It is necessary to gradually scale up each job.
    Therefore, we allow a running job to request to scale up its batch size after each training epoch, by doubling the batch size limit as $R'=2R$.
    \item \textit{Scale-down:}
    A running job with long elapsed time will need to be scaled down to prevent the Convoy Effect.
    We penalize the job with
    \begin{align}
        \nonumber R' = \lceil \frac{2R}{\lceil \sigma \cdot T_{processed}+1 \rceil} \rceil,
    \end{align}
    where $T_{processed}$ denotes the executed time of the job and $\sigma$ denotes a fixed factor.
    We suggest $\sigma=\lambda$ where $\lambda$ denotes the average job arrival rate, in order to penalize jobs that are longer than the average arrival time interval $1/\lambda$.
 \end{itemize}

% \subfile{section/implementation.tex}

\section{Evaluation}
\label{sec:eval}

\subsection{Setups}

\textbf{Testbed:}
We evaluate ONES on TACC's Longhorn supercomputers \cite{longhorn}.
Our testbed cluster consists of 16 GPU servers connected by Mellanox EDR Infiniband network.
Each server has 2 20-core IBM Power 9 CPUs, 256GB memory and 4 NVIDIA V100 GPUs, so there are 64 GPUs in total.
We store datasets, model checkpoints and training logs in a Hadoop Distributed File System (HDFS) via 1Gbps Ethernet.

\textbf{Trace-driven workload:}
% We follow the job distributions in real production traces from Microsoft \cite{jeon2019analysis} to generate our workload for evaluation.
% In the workload we include the jobs that are popularly used in the experiments of prior works, as listed in Table \ref{tab:jobs}.
% However, Some of the jobs (e.g. training ImageNet and WMT14) can take weeks long to complete the training.
% We downscale the dataset sizes (ImageNet from 1.28m images, 1000 classes to 64k images, 200 classes; WMT14 from 1m sentences to 40k sentences) so that the experiments can be limited in a reasonable amount of time.
% Thereby, the training time of downscaled jobs using the batch size recommended in corresponding references ranges from 1 minute to 6 hours.
% We then generate the workload with the proportional job length distribution and average arrival time interval to those in the Microsoft traces.
We generate our custom traces with typical DL tasks that are popularly used in the experiments of prior works \cite{peng2018optimus, xiao2018gandiva, gu2019tiresias}, as listed in \autoref{tab:jobs}.
These tasks include both computer vision (CV) and natural language processing (NLP) models and datasets, including AlexNet \cite{dean2012large}, ResNet \cite{he2016deep}, VGG \cite{simonyan2014very}, GoogleNet \cite{szegedy2015going}, Inception \cite{szegedy2016rethinking} and BERT \cite{devlin2018bert}.
We implement the tasks with distributed PyTorch.
To generate workloads with sufficient diversity, we train the jobs with different dataset sizes instead of the original sizes (e.g. using a subset of 10k and 20k images from the original ImageNet dataset).
Consequently, the training time of some long jobs is reduced, and all jobs can basically finish within 2 hours.
As listed in \autoref{tab:jobs}, the number of different workloads is $4\times6+3\times5+4+1+6=50$.
\B{Normally ONES ends a job when there is no increase in its validation accuracy for several consecutive epochs. In our implementation and experiments, we set a target validation accuracy for each job in \autoref{tab:jobs}, and stop the training when there are already 10 consecutive epochs that exceed the target accuracy. The reason is to make sure that, in comparison between ONES and baseline schedulers, different schedulers can provide similar convergence for each job.}

\begin{table}[t]
    \centering
    \begin{tabular}{c|c|c|c|c}
        \toprule
        Task & Dataset & Model & Dataset Size & \# Classes\\
        \hline
        \multirow{5}{*}{CV} & \multirow{4}{*}{ImageNet} & AlexNet & \multirow{4}{*}{\makecell[c]{10k, 12k, \\14k, ..., 20k}} & \multirow{4}{*}{\makecell[c]{10, 12, \\14, ..., 20}}\\
        \cline{3-3}
        & & ResNet50 & \\
        \cline{3-3}
        & & VGG16 & \\
        \cline{3-3}
        & & InceptionV3  & \\
        \cline{2-5}
        & \multirow{3}{*}{CIFAR10} & ResNet18 & \multirow{3}{*}{\makecell[c]{20k, 25k, 30k,\\35k, 40k}} & \multirow{3}{*}{10}\\
        \cline{3-3}
        & & VGG16  & \\
        \cline{3-3}
        & & GoogleNet & \\
        \hline
        \multirow{5}{*}{NLP} & \multirow{2}{*}{COLA} & \multirow{5}{*}{\makecell[c]{pre-trained\\BERT}} & \multirow{2}{*}{\makecell[c]{5k, 6k, \\7k, 8k}} & \multirow{2}{*}{2}\\
        & & & &\\
        \cline{2-2}\cline{4-5}
        & MRPC & & 3.6k & 2\\
        \cline{2-2}\cline{4-5}
        & \multirow{2}{*}{SST-2} & & \multirow{2}{*}{\makecell[c]{10k, 12k, \\14k, ..., 20k}} & \multirow{2}{*}{2}\\
        & & &\\
        \bottomrule
    \end{tabular}
    \caption{Workloads in our evaluation trace.}
    \label{tab:jobs}
\end{table}

% \begin{table*}[!t]
%     \centering
%     \begin{tabular}{c|c|c|c|c}
%         \toprule
%         Model & Task & Dataset & Dataset size & Target\\
%         \hline
%         AlexNet \cite{krizhevsky2012imagenet} & Image classification & ImageNet \cite{imagenet} & 64k & Top-1 Acc: 86\% \\
%         \hline
%         ResNet50 \cite{he2016deep} & Image classification & ImageNet & 64k & Top-1 Acc: 86\% \\
%         \hline
%         VGG16 \cite{simonyan2014very} & Image classification & ImageNet & 64k & Top-1 Acc: 86\% \\
%         \hline
%         ResNet152 \cite{he2016deep} & Image classification & CIFAR10 \cite{cifar} & 50k & Top-1 Acc: 97\% \\
%         \hline
%         InceptionV3 \cite{szegedy2016rethinking} & Image classification & CIFAR10 & 50k & Top-1 Acc: 97\% \\
%         \hline
%         BERT \cite{devlin2018bert} & Sentence similarity & MRPC \cite{dolan2005automatically} & 3.7k & Acc: 84\% \\
%         \hline
%         Transformer \cite{vaswani2017attention} & Machine translation & WMT14 \cite{wmt14} & 40k & BLEU: 27 \\
%         \bottomrule
%     \end{tabular}
%     \caption{Jobs}
%     \label{tab:jobs}
% \end{table*}

\textbf{Baselines:}
We compare the performance of ONES against the following state-of-the-art DL schedulers \B{that also aim to minimize JCT}.
\begin{itemize}
    \item Deep Reinforcement Learning (\textit{DRL}):
    We adopt the basic scheduler design in \cite{gong2019chic} but modify its action space because we use the All-reduce architecture for distributed training instead of parameter servers.
    The scheduler trains its scheduling policy based on DRL for purpose of minimizing JCT. It can dynamically determine the size of each job. Only one job can be rescheduled at each time.
    \item \textit{Tiresias} \cite{gu2019tiresias}:
    This scheduler maintains the waiting jobs in a multi-level priority queue according their attained service to reduce JCT. It only allows fixed job size but does not depend on any prediction.
    \item \textit{Optimus} \cite{peng2018optimus}:
    This scheduler can dynamically adjust the size of each job, and uses a greedy strategy to allocate resources based on the predicted remaining processing time of each job so as to reduce average JCT.
    % \item \textit{Gandiva} \cite{xiao2018gandiva}:
    % This scheduler uses a time-sharing mechanism to pack multiple jobs on the same GPU so as to reduce latency and improve cluster utilization. It can grow the size of some job when there is idle resource. Its scheduling policy is static without any prediction.
\end{itemize}

\B{As shown in \autoref{tab:baseline}, Optimus and Tiresias adopt greedy scheduling strategies, while ONES and DRL are dynamic because they are based on evolutionary search and deep reinforcement learning respectively.
Different to the baseline schedulers, ONES is the only one that is able to allow elastic batch sizes.
In contrast, Tiresias even does not allow elastic job size so that the number of GPUs of each job is fixed after submission.
DRL cannot allow preemption so that it does not stop any job during its training until it is completed, which to some extent affects the scheduling efficiency because DL jobs are usually long, and appropriate job preemption can be beneficial to reduce average JCT.}

\textbf{Metrics:}
We use average job completion time (JCT) as the foremost metric of scheduling performance.
The JCT of a job is the time period between its submission and completion.
To clearly present the performance of ONES, we also provide box plots and cumulative frequency (CF) curves to illustrate the distribution over the JCT of all the jobs.

% We also evaluate different components of JCT, including the execution time and queuing time of each job, so as to further examine the efficiency of our scheduler. 

\begin{table}[t]
    \centering
    \resizebox{\linewidth}{!}{
    \begin{tabular}{c|c|c|c|c}
        \toprule
        Scheduler & \makecell[c]{Greedy/Dynamic\\Strategy} & \makecell[c]{Allow\\Preemption} & \makecell[c]{Elastic\\Job Size} & \makecell[c]{Elastic\\Batch Size} \\
        \hline
        ONES & Dynamic & Y & Y & Y\\
        \hline
        DRL & Dynamic & N & Y & N \\
        \hline
        Tiresias & Greedy & Y & N & N \\
        \hline
        Optimus & Greedy & Y & Y & N \\
        \bottomrule
    \end{tabular}}
    \caption{\B{Comparison of ONES and the state-of-the-art DL schedulers.}}
    \label{tab:baseline}
\end{table}

\subsection{Performance}

\begin{figure*}[t]
    \centering
    \begin{subfigure}[t]{0.32\textwidth}
        \centering
        \includegraphics[width=\linewidth]{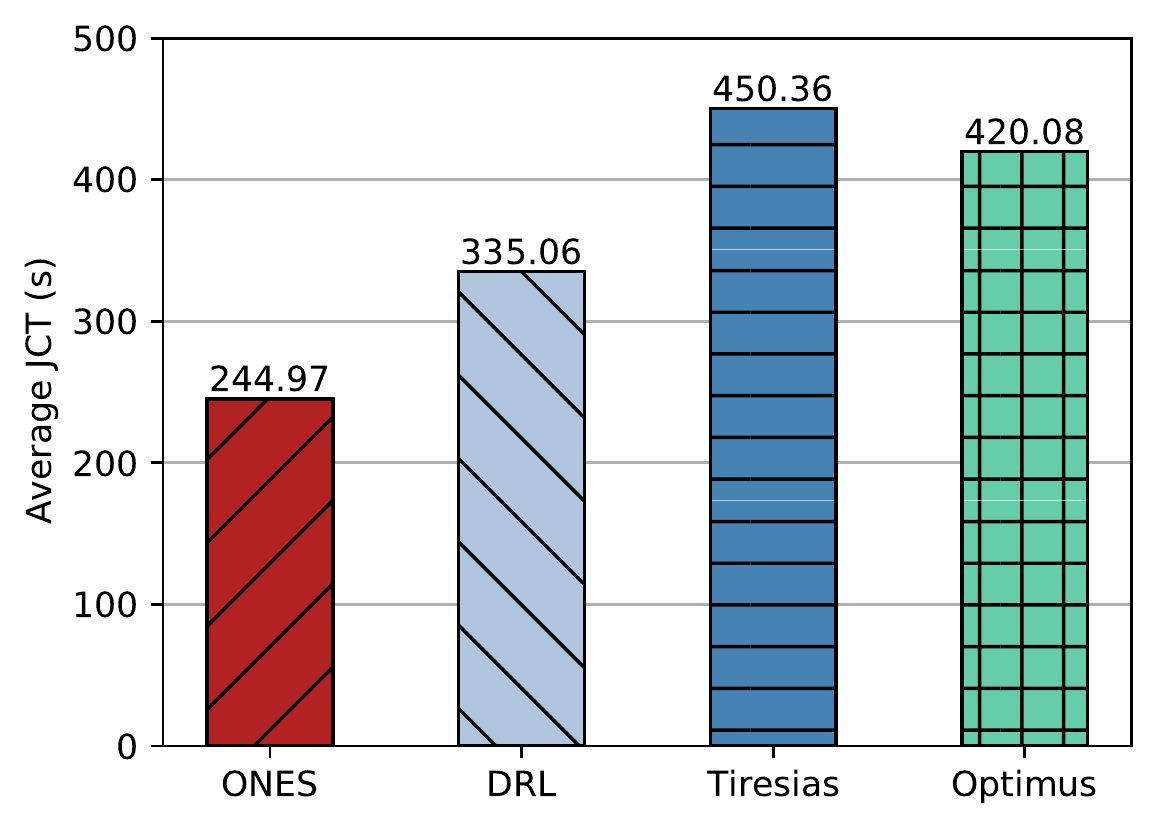}
        \caption{Average completion time}
        \label{fig:jct}
    \end{subfigure}
    \begin{subfigure}[t]{0.32\textwidth}
        \centering
        \includegraphics[width=\linewidth]{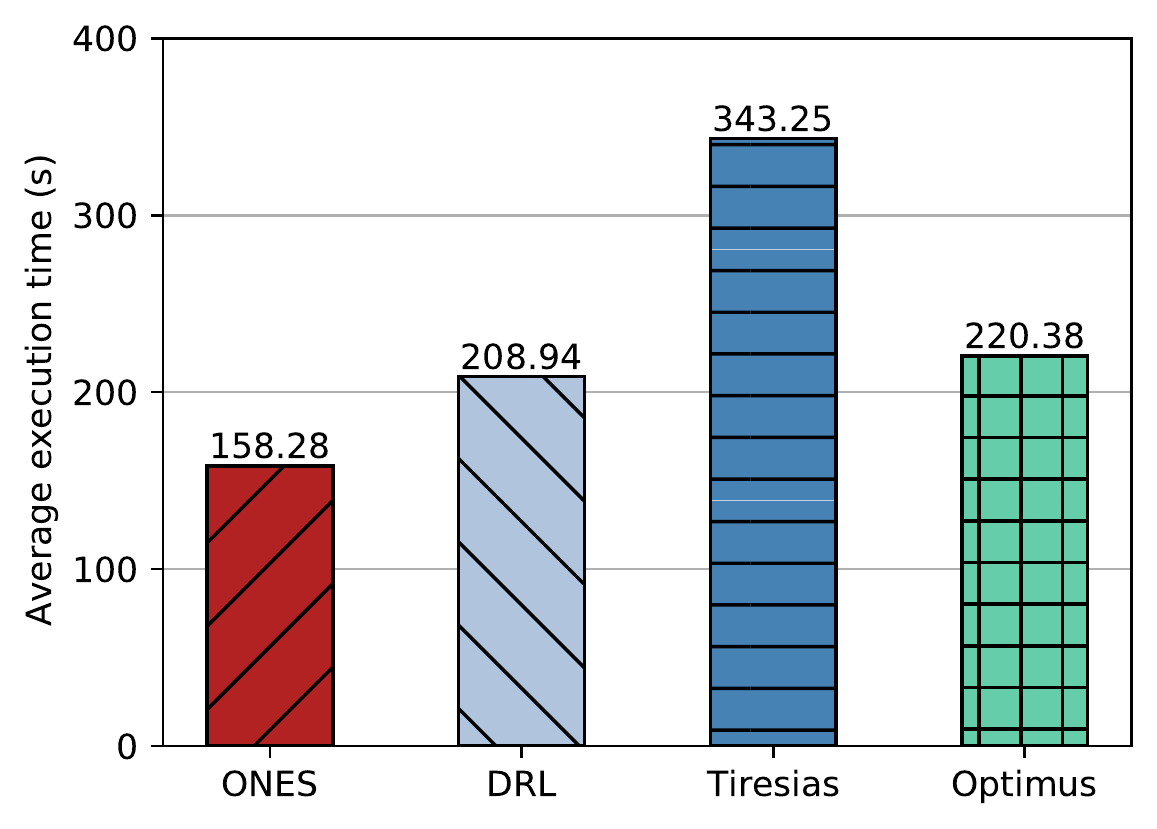}
        \caption{Average execution time}
        \label{fig:exec}
    \end{subfigure}
    \begin{subfigure}[t]{0.32\textwidth}
        \centering
        \includegraphics[width=\linewidth]{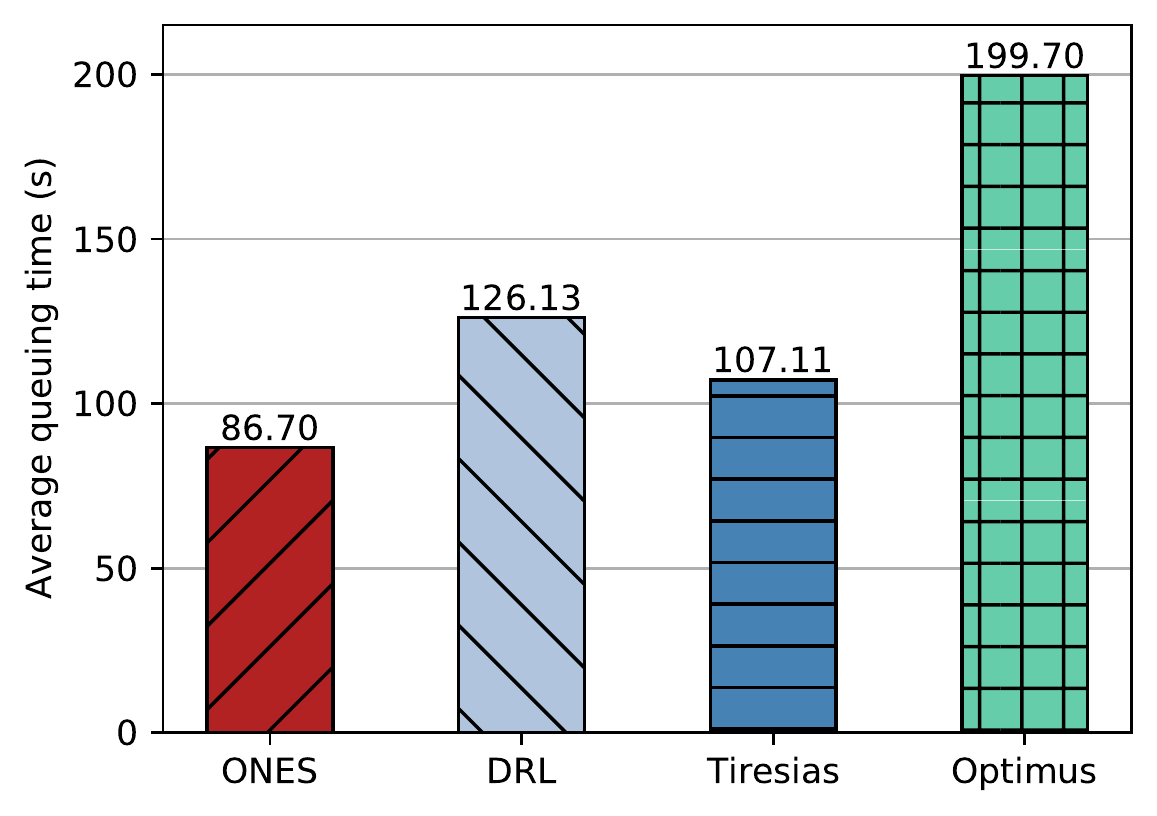}
        \caption{Average queuing time}
        \label{fig:delay}
    \end{subfigure}
    \begin{subfigure}[t]{0.32\textwidth}
        \centering
        \includegraphics[width=\linewidth]{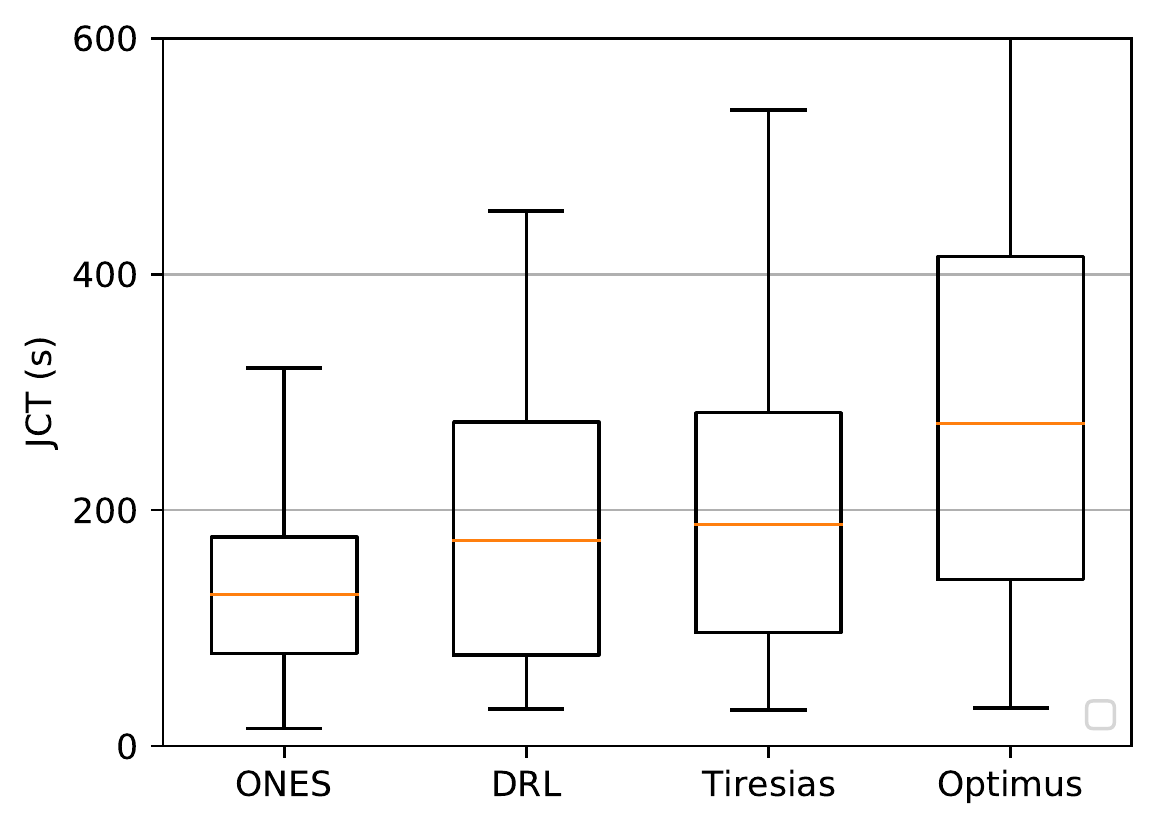}
        \caption{Completion time distributions}
        \label{fig:jct-box}
    \end{subfigure}
    \begin{subfigure}[t]{0.32\textwidth}
        \centering
        \includegraphics[width=\linewidth]{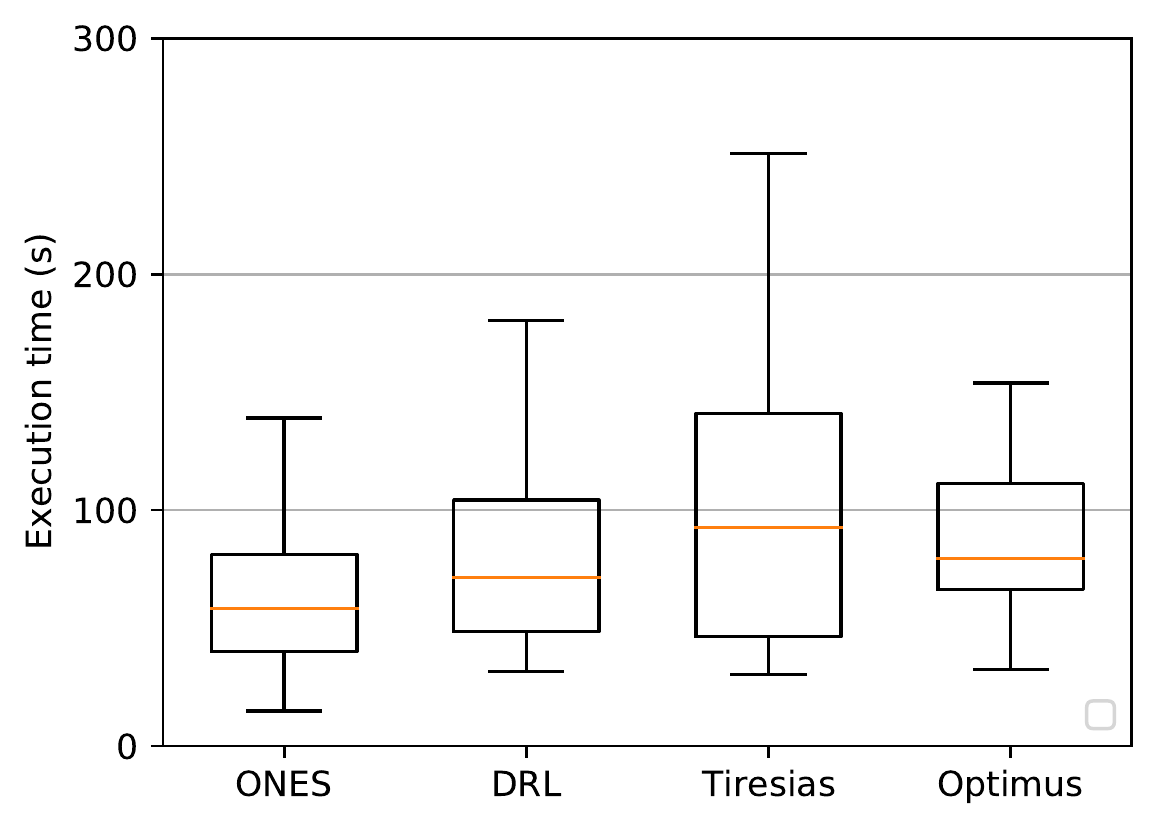}
        \caption{Execution time distributions}
        \label{fig:exec-box}
    \end{subfigure}
    \begin{subfigure}[t]{0.32\textwidth}
        \centering
        \includegraphics[width=\linewidth]{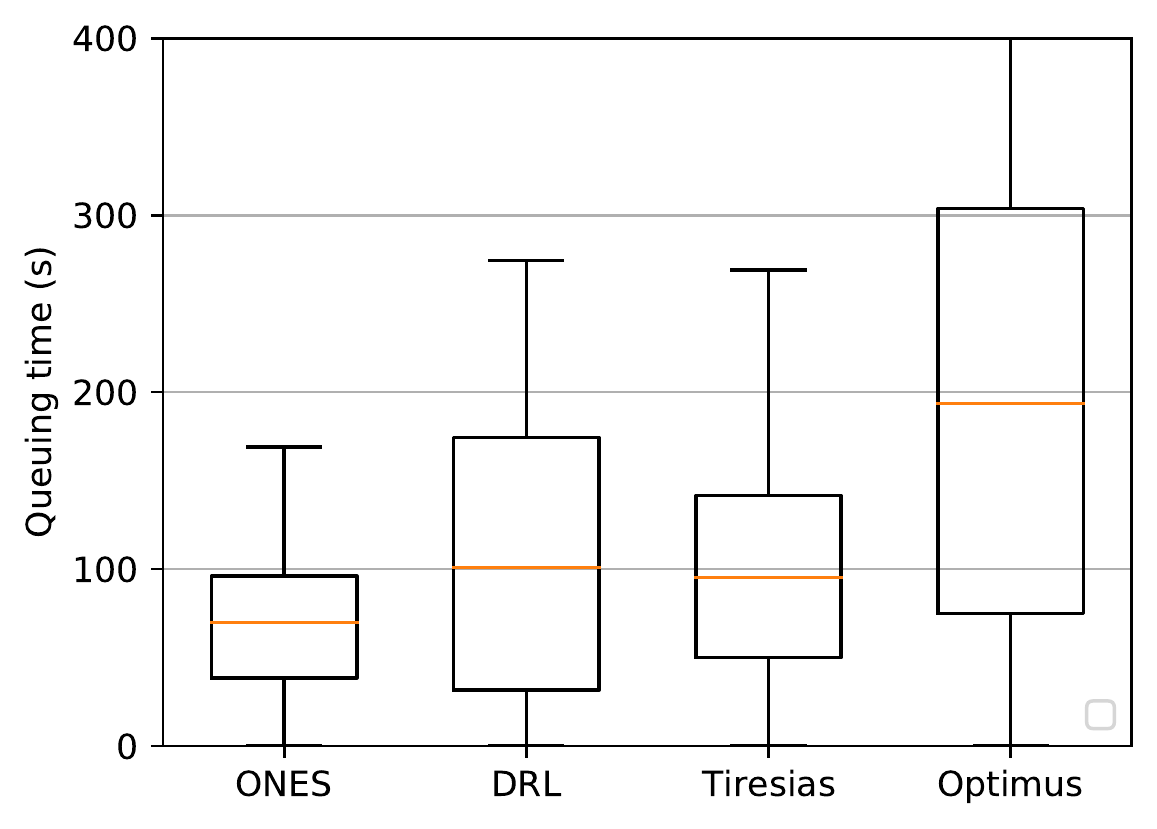}
        \caption{Queuing time distributions}
        \label{fig:delay-box}
    \end{subfigure}
    \begin{subfigure}[t]{0.32\textwidth}
        \centering
        \includegraphics[width=\linewidth]{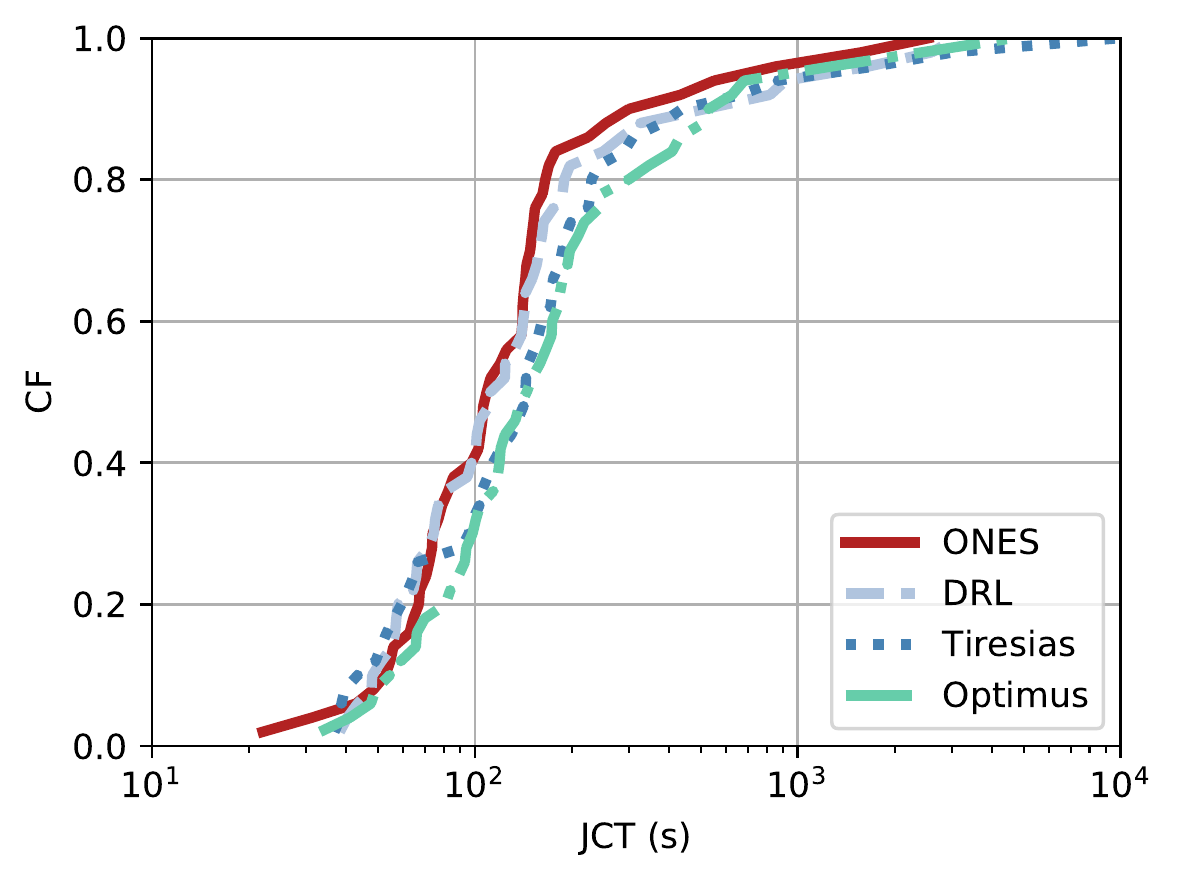}
        \caption{Completion time cumulative frequency curves}
        \label{fig:jct-cdf}
    \end{subfigure}
    \begin{subfigure}[t]{0.32\textwidth}
        \centering
        \includegraphics[width=\linewidth]{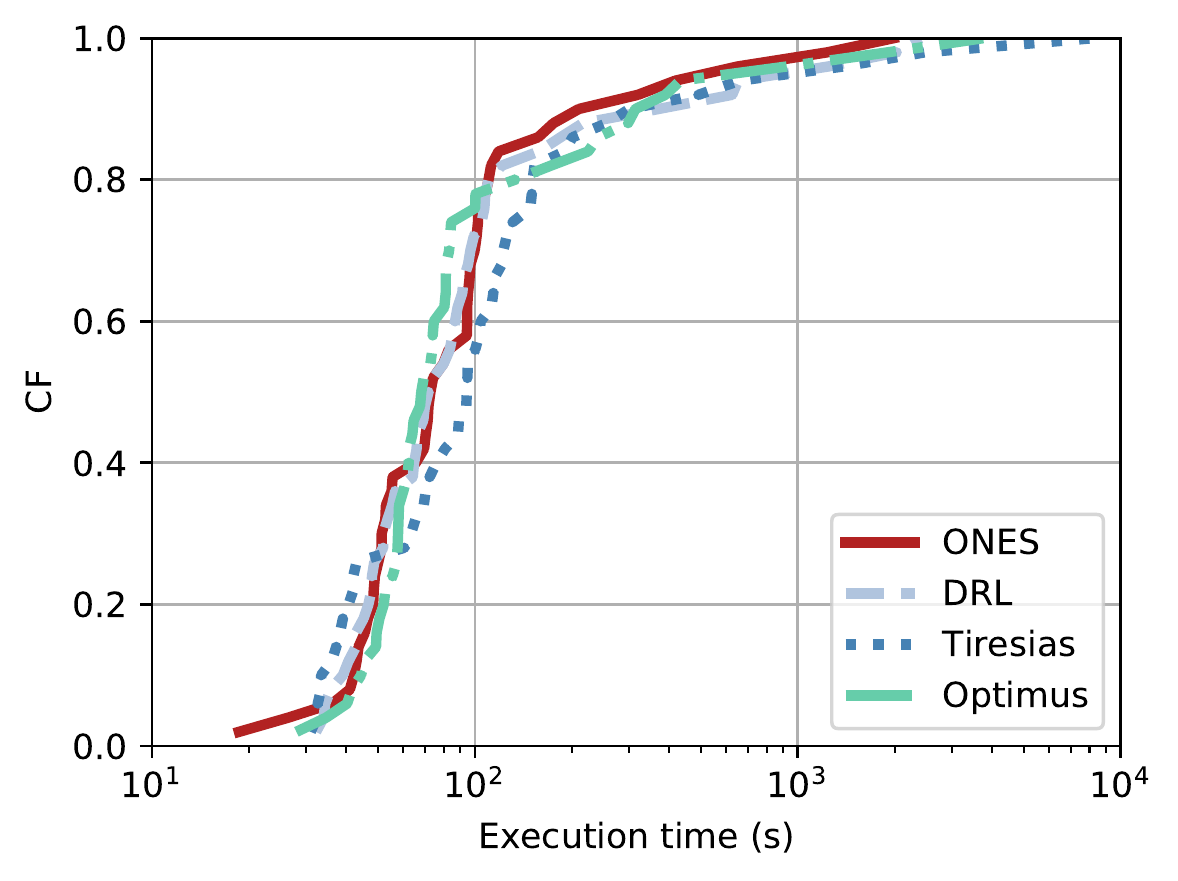}
        \caption{Execution time cumulative frequency curves}
        \label{fig:exec-cdf}
    \end{subfigure}
    \begin{subfigure}[t]{0.32\textwidth}
        \centering
        \includegraphics[width=\linewidth]{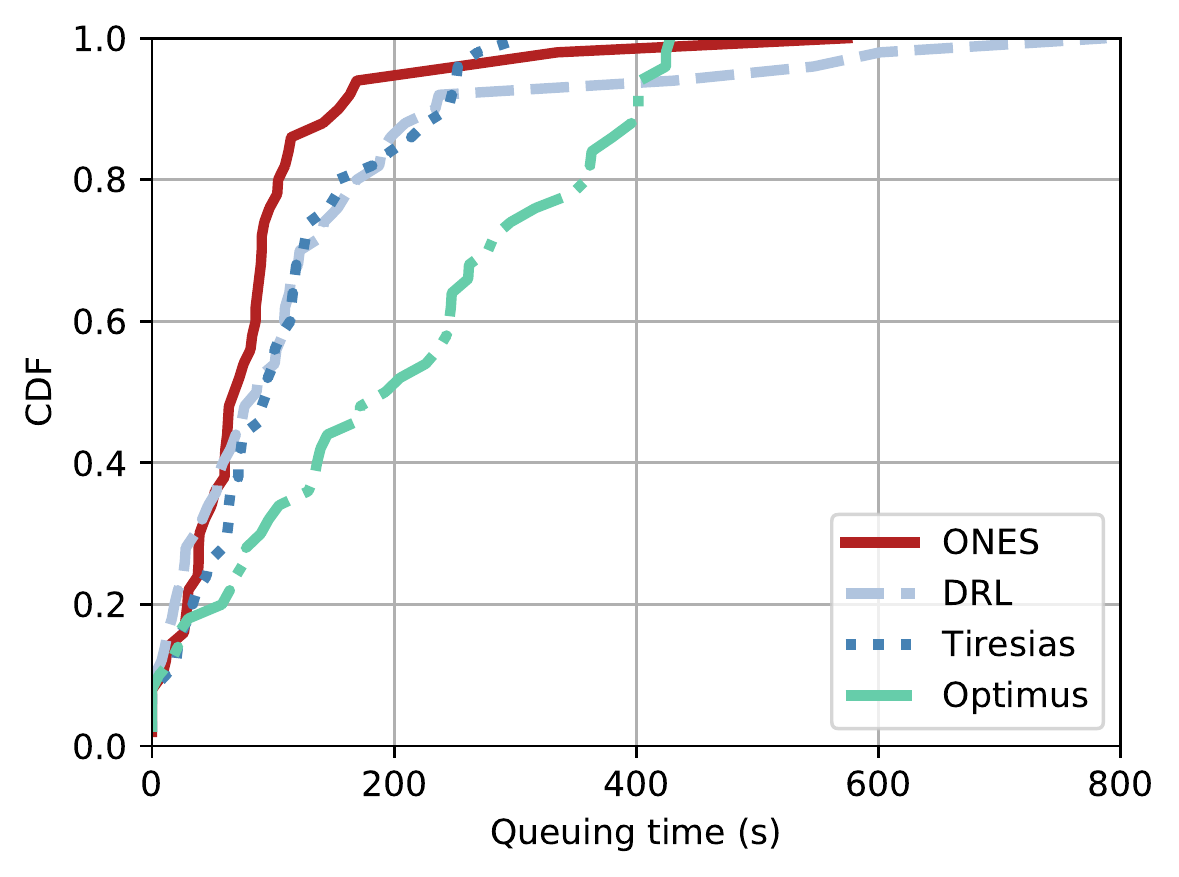}
        \caption{Queuing time cumulative frequency curves.}
        \label{fig:delay-cdf}
    \end{subfigure}
    \caption{Comparisons of the scheduling performance.}
\end{figure*}

\begin{table}[t]
    \centering
    \begin{tabular}{c|c|c}
        \toprule
        & \makecell[c]{$p$ value\\(two-sided test)} & \makecell[c]{$p$ value \\(one-sided \\negative test)}\\
        \hline
        vs. DRL & 5.17e-4 & 0.99974 \\
        \hline
        vs. Tiresias & 4.53e-8 & 0.99999 \\
        \hline
        vs. Optimus & 7.55e-10 & 0.99999 \\
        \bottomrule
    \end{tabular}
    \caption{\B{Significance tests using Wilcoxon test}}
    \label{tab:wilcoxon}
\end{table}

\textbf{JCT comparison:}
\autoref{fig:jct} shows the results of average JCT.
We see that ONES can reduce the average JCT by 26.9\%, 45.6\% and 41.7\%, compared to DRL, Tiresias \cite{gu2019tiresias} and Optimus \cite{peng2018optimus}, respectively.
ONES can achieve the smallest average JCT of 244.97s.
To see more details, \autoref{fig:jct-box} and \ref{fig:jct-cdf} shows the JCT distributions.
We see that ONES is effective to reduce JCT for especially long jobs, as the fraction of jobs completed within 200s is 86\%, while the baseline schedulers can only achieve about 60-80\%.
From the distribution, we observe that ONES is especially effective in reducing the completion time of slow jobs, which greatly contributes to the reduction in average completion time.
\B{\autoref{tab:wilcoxon} shows comparisons of ONES against baseline schedulers based on significance tests.
We use non-parametric Wilcoxon tests \cite{wilcoxon1992individual} to present statistical analysis of each job's JCT and evaluate whether ONES outperforms the baselines.
The two-sided tests evaluate a hypothesis that the results of ONES and the other scheduler are equivalent.
Since the tests give $p$ values (i.e. confidence) that are much smaller 0.05, we can reject their hypotheses.
Next, The one-sided negative tests evaluate a hypothesis that the results of ONES are smaller than the other scheduler.
We can accept the hypotheses and consider that ONES outperforms the baseline schedulers, because the tests give very high $p$ values.}

To explain the performance improvement of ONES, we then present further evaluations by breaking down the completion time of each job into execution time and queuing time.

% \begin{figure}[t]
%     \centering
%     \begin{minipage}[t]{0.45\linewidth}
%         \centering
%         \includegraphics[width=\linewidth]{figure/exec.pdf}
%         \caption{Comparison\\ of average execution\\ time.}
%         \label{fig:exec}
%     \end{minipage}
%     \begin{minipage}[t]{0.45\linewidth}
%         \centering
%         \includegraphics[width=0.9\linewidth]{figure/exec-cdf.pdf}
%         \caption{Comparison of execution time distributions.}
%         \label{fig:exec-cdf}
%     \end{minipage}
% \end{figure}

\textbf{Training faster:}
By dynamically scaling the batch size and GPU resources of each job, ONES makes its best effort to improve resource utilization and reduce training time.
\autoref{fig:exec}, \ref{fig:exec-box} and \autoref{fig:exec-cdf} present the results of ONES and show its efficiency in reducing execution time, compared to the baselines.
While the Tiresias scheduler is not able to adjust the resource size for each job, ONES can sufficiently utilize the entire cluster and reduce 53.9\% execution time.
Besides, ONES achieves smaller execution time than both DRL and Optimus, because ONES can eliminate the performance bottleneck between the increasing number of GPUs and the decreasing training speed per GPU.
As a result, for jobs with a relatively larger batch size, ONES can achieve better performance than the DRL and Optimus.

The distribution show that for fast jobs ONES has similar performance to the others, whereas ONES can obviously shorten the execution time of slow jobs.
Basically all the jobs is completed under 2k seconds.
However other schedulers all have jobs with longer execution time up 10k seconds.

% \begin{figure}[t]
%     \centering
%     \begin{minipage}[t]{0.45\linewidth}
%         \centering
%         \includegraphics[width=\linewidth]{figure/delay.pdf}
%         \caption{Comparison\\ of average queuing\\ time.}
%         \label{fig:delay}
%     \end{minipage}
%     \begin{minipage}[t]{0.45\linewidth}
%         \centering
%         \includegraphics[width=0.9\linewidth]{figure/delay-cdf.pdf}
%         \caption{Comparison of queuing time distributions.}
%         \label{fig:delay-cdf}
%     \end{minipage}
% \end{figure}

\begin{figure}[t]
    \centering
    \includegraphics[width=\linewidth]{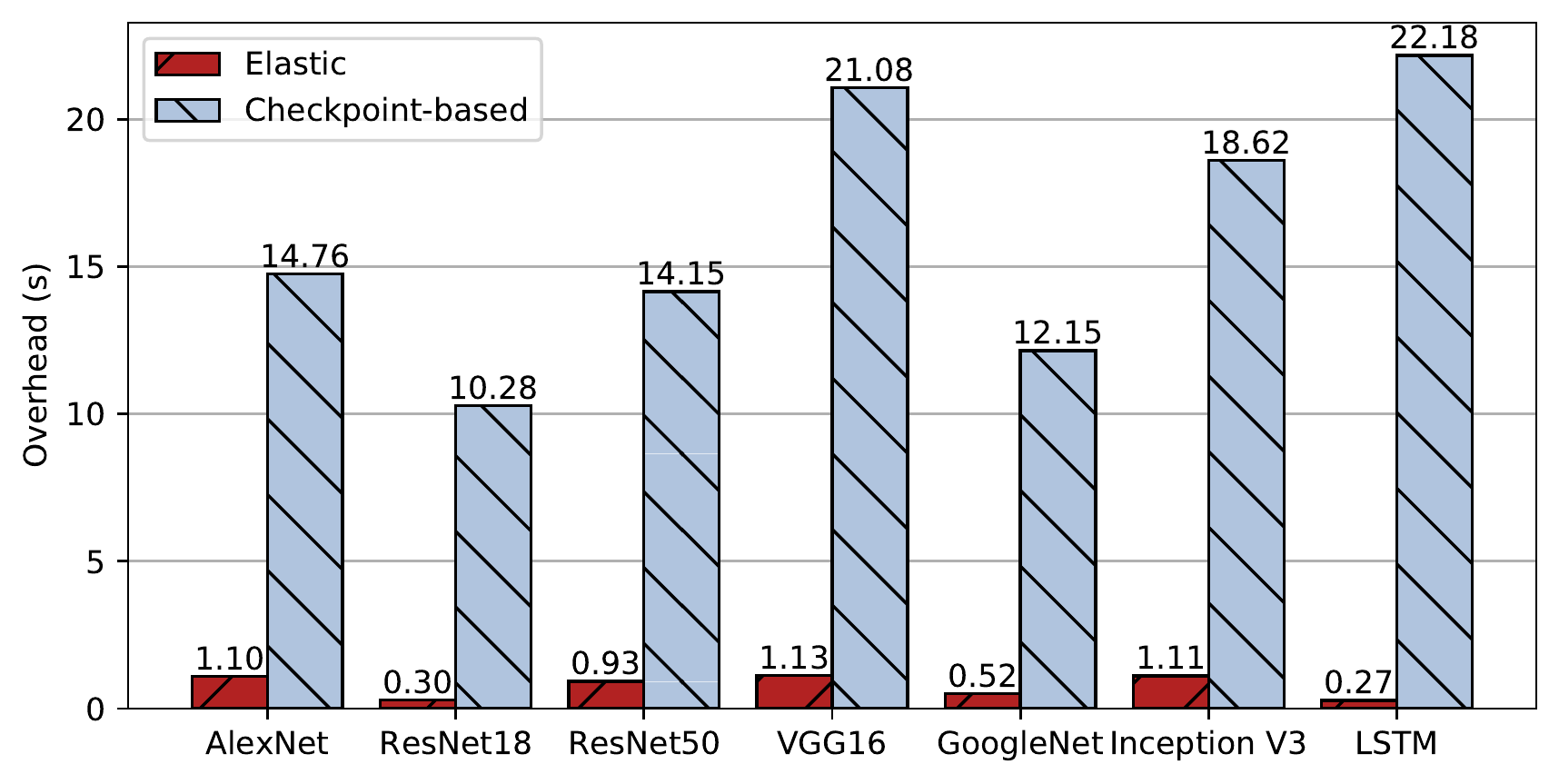}
    \caption{Comparison between the overheads of elastic batch size scaling and checkpoint-based migration.}
    \label{fig:overhead}
\end{figure}

\textbf{Waiting less:}
\autoref{fig:delay}, \ref{fig:exec-box} and \ref{fig:delay-cdf} show the queuing time performance of ONES.
The online evolutionary algorithm (\S\ref{sec:online}) can help our scheduler achieve the smallest queuing time, compared to the strategies used by the baselines.
We find that ONES can reach the smallest average queuing time, though its maximum queuing time may be longer.
In overall, ONES improves the DRL scheduler by 31.3\%, Tiresias by 19.1\%, and Optimus by 56.6\%.
However, the queuing time performance of Optimus is a special case.
The Optimus scheduler runs in a periodical pattern.
Suppose jobs arrive at completely random time, the expected average queuing time is supposed to be half of the scheduling interval.
When smaller scheduling interval leads to smaller average queuing time, it also results in higher overhead due to frequent rescheduling.
In our experiments, we adopt the same scheduling interval of 10 minutes as in the Optimus paper \cite{peng2018optimus}.

\subsection{Overhead Analysis}

The major overhead of ONES comes from job size scaling.
In this section, we compare the performance between our elastic batch size scaling and checkpoint-based migration for training different models.
Checkpoint-based migration is a common approach used by DL jobs today.
It first stops the training and saves a job status into checkpoint files, and then resumes the job from the checkpoint with new configurations.
\autoref{fig:overhead} illustrates the result of the scaling overheads of different models implemented in PyTorch.
The result shows that the overhead of elastic batch size scaling is basically around 1 second, while that of checkpoint-based migration can be greater than 20 seconds.
Checkpoint-based approach mainly wastes time on preparing data and loading model to GPU devices.
In contrast, they are not necessary in our approach.
Therefore, our approach is able to significantly reduce the overhead.

\subsection{Scalability Analysis} 

Lastly, we examine the scheduling scalability by evaluating the average JCT performance under different cluster capacities.
In this evaluation, we scale the cluster capacity from 16 GPUs to 64 GPUs.

\begin{figure}[t]
    \centering
    \includegraphics[width=0.99\linewidth]{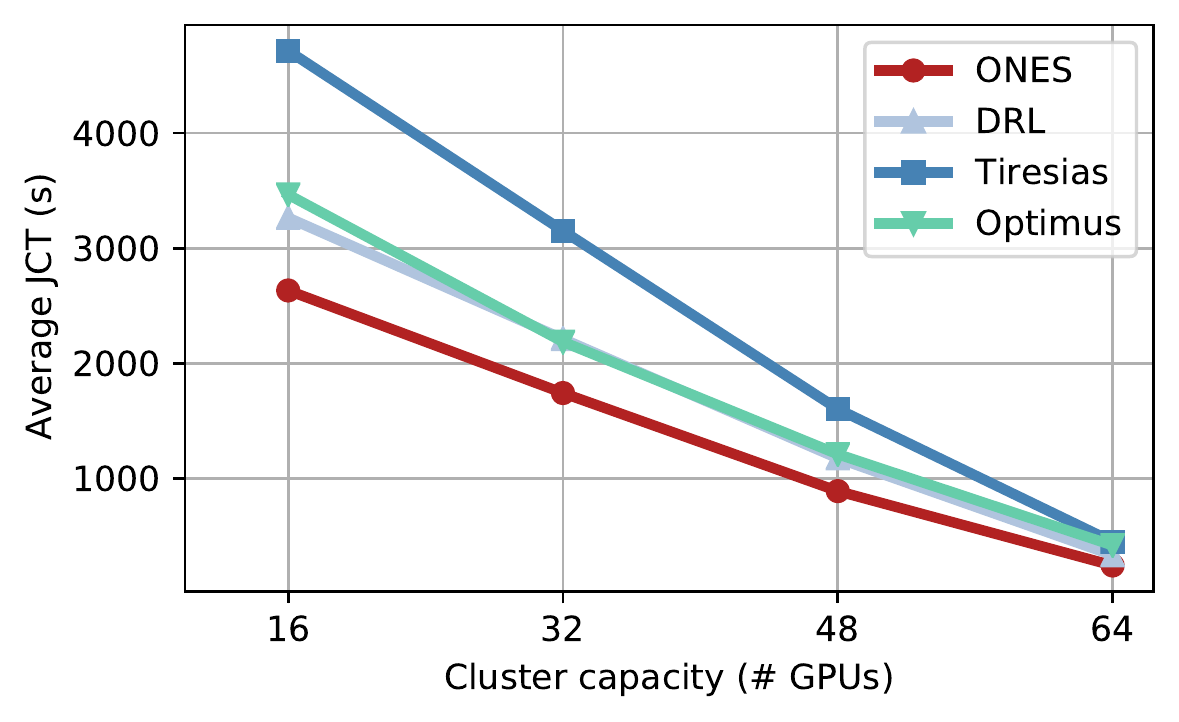}
    \caption{Comparison of scheduling scalability.}
    \label{fig:performance-scaling}
\end{figure}

\begin{figure}[t]
    \centering
    \includegraphics[width=\linewidth]{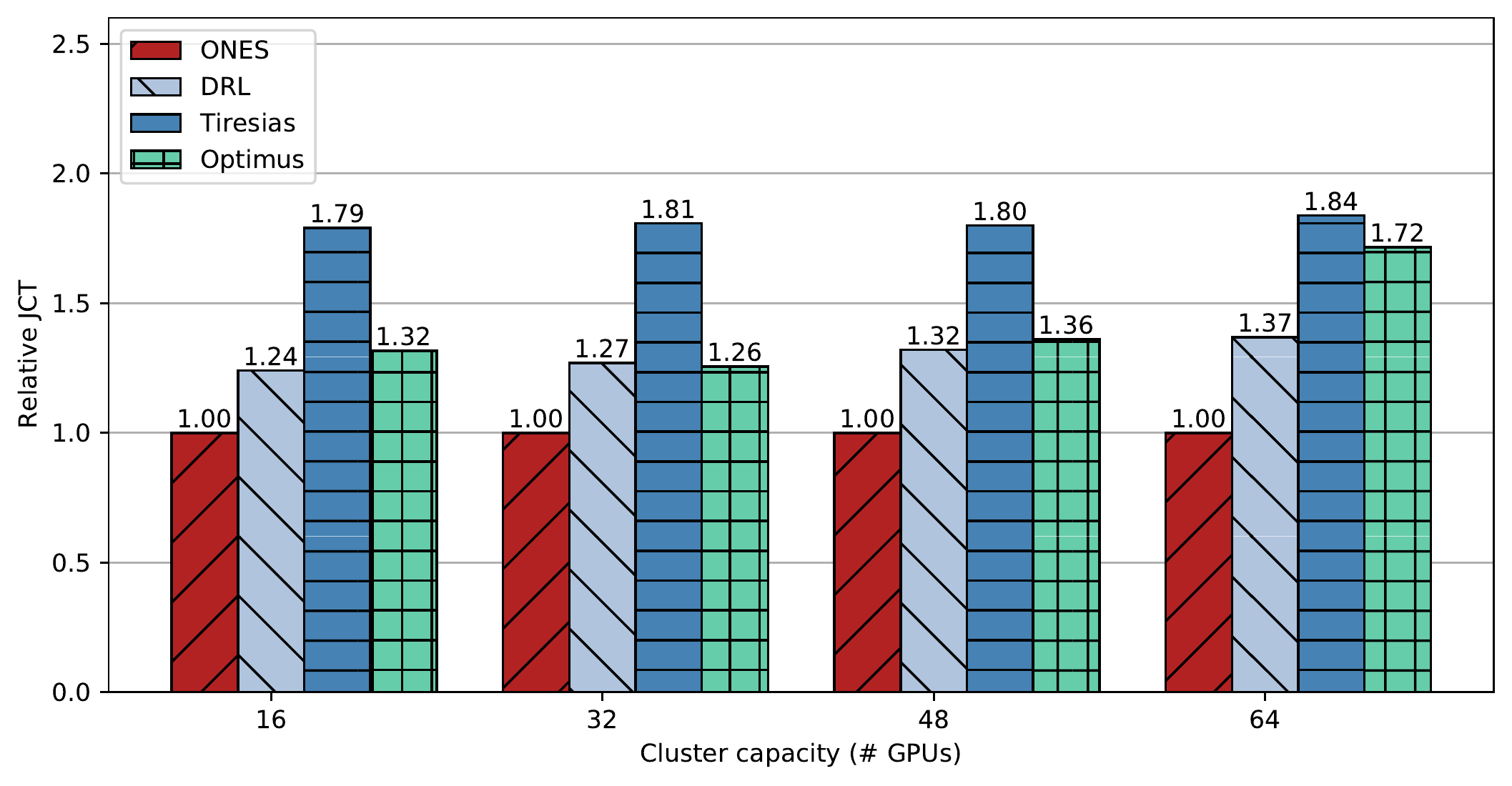}
    \caption{Scalability of improvements in JCT.}
    \label{fig:improvement-scaling}
\end{figure}

The results are presented in \autoref{fig:performance-scaling} and \autoref{fig:improvement-scaling}.
We see that all the schedulers are supposed to show similar trends.
By increasing the number of GPUs, the average JCT will be almost linearly reduced.
Especially the average queuing time will decrease linearly.
Besides, we notice that by increasing the number of GPUs from 16 to 64, the improvement of ONES also increases from 19\%/44\%/24\% to 27\%/46\%/42\%, compared to DRL, Tiresias and Optimus respectively.
with more GPUs (for example, with 64 GPUs in the figures), ONES may have a larger improvement.
In other words, compared with the baselines, ONES can make the most efficient use of free GPU resources.

\section{Related Works}
\label{sec:review}

Many prior works develop a variety of analytic performance models to optimize their resource scheduling algorithms based on different objectives.
For example, SLAQ \cite{zhang2017slaq} adopts an online fitting model to estimate the training quality for future steps of each job by collecting the quality and resource usage information from concurrent jobs.
Accordingly, it schedules the jobs to machines for maximizing the overall improvement in the training quality of all the jobs.
Similarly, to minimize average JCT, Optimus \cite{peng2018optimus} periodically adjusts the amount of resources for the DDL jobs with PS architecture.
For each schedule, it builds a resource-speed model for each job, and then continuously adds a worker or a parameter server to the job with the maximum decrease in estimated JCT until the cluster is full.
It allocates at least one worker and one parameter server to each job for fairness, whereas such policy may result in inefficiency when the cluster is over-subscribed.
% Besides, Optimus may not be efficient from the online perspective, since it has to adjust the schedule for the entire cluster with a constant rescheduling interval (e.g. every 10 minutes).
OASiS \cite{oasis} gives the analysis of the performance of asynchronous training.
It also presents a scheduler based on the greedy policies for solving an integer programming problem that aims to maximize the overall system utilization.
% However, the use case of OASiS is limited since asynchronous training is not commonly used by DL applications due to stale gradients.

Some other DL schedulers put many efforts in developing job prioritization and placement policies.
Gandiva \cite{xiao2018gandiva} uses a time-slicing mechanism on GPU resources across multiple jobs to reduce the latency and improve cluster utilization.
It works in an introspective way that continuously packs and migrates job placement to optimize the locality performance.
Tiresias \cite{gu2019tiresias} attempts to reduce average JCT under the assumption that job lengths are hard to be measured in advance, which applies to our work as well.
It proposes discretized priority queues for scheduling, based on attained service time of each job. But it cannot dynamically adjust the number of GPUs allocated to each job.
Philly \cite{jeon2019analysis} is designed based on the analysis of the trace from Microsoft DL cluster. The analysis indicates the affect of the locality of DL jobs on queuing delay and GPU utilization, and then gives implications that schedulers should consider prioritizing locality, mitigating interference, as well as improving failure handling.

Apart from greedy scheduling algorithms, there is a recent trend in learning scheduling policy based on deep reinforcement learning \cite{mao2019learning, gong2019chic, peng2019dl2}.
On one hand, the performance of such schedulers heavily relies on the training traces because they assume that the distribution of scheduling policy will not change over time.
In practice, whether the policy distribution can always work should be questionable.
On the other hand, the DRL agent of such schedulers produces one action at each time, while each action usually works with only one job, so as to avoid oversized action space.
Besides, the iteration speed of DRL schedulers is apparently slower than evolutionary search.

\section{Conclusion}
\label{sec:conclude}

The focus of this paper is to study the resource scheduling problem for deep learning workloads in a shared GPU cluster.
We aim to optimize the overall training performance by allocating an appropriate batch size to each job in real-time.
% In this paper, we investigate the performance of distributed training jobs and propose an online evolutionary scheduler, which manages an elastic batch size of each job to improve utilization and minimize the average job completion time.
% The evaluations show that the scheduler is efficient in minimizing JCT with smaller scheduling overheads, compared with the prior state-of-the-art schedulers.
In this paper, we propose ONES, which is the first to schedule the elastic batch sizes instead of the number of GPUs for DL jobs.
The scheduler dynamically manages the batch sizes through an online evolutionary search algorithm and efficient batch size scaling mechanism.
As a result, ONES is able to reduce the average JCT by up to 45.6\%, compared to state-of-the-art methods.

\section{Acknowledgments}
We thank CSCS (Swiss National Supercomputing Centre) for supporting our project to get access to the Piz Daint supercomputer. We thank TACC (Texas Advanced Computing Center) for supporting our project to get access to the Longhorn supercomputer and the Frontera supercomputer. We thank LuxProvide (Luxembourg national supercomputer HPC organization) for supporting our project to get access to the MeluXina supercomputer.

\bibliographystyle{ACM-Reference-Format}
\bibliography{sample-base}

% \appendix

% \subfile{section/appendix.tex}









\end{document}